\documentclass[10pt,a4paper,notitlepage,prd,showpacs,preprintnumbers,superscriptaddress,nofootinbib]{revtex4-1}
\usepackage{amsfonts,amsmath,amssymb,bm,tensor}
\usepackage{comment}
\usepackage{ifpdf}
\usepackage{slashed}
\usepackage{color}
\usepackage[mathscr]{eucal}
\usepackage[utf8]{inputenc}
 
\usepackage{cancel}

\ifpdf
  \usepackage{graphicx}     %   usepackage without driver option
  \usepackage[bookmarksopen,colorlinks=true,linkcolor=bblue,citecolor=bblue,urlcolor=ppink]{hyperref}
\else     % For (p)LaTeX + dvipdfmx
  \usepackage[dvipdfmx]{graphicx}     %   usepackage with driver option
  \usepackage[dvipdfmx,bookmarksopen,colorlinks=true,linkcolor=bblue,citecolor=ppink,urlcolor=darkred]{hyperref}
\fi

\definecolor{red}{rgb}{1,0,0}
\definecolor{darkred}{rgb}{0.6,0,0}
\definecolor{darkgreen}{rgb}{0.992447,0.623778,0.034597}
\definecolor{ppink}{rgb}{1,0.4,0.4}
\definecolor{bblue}{rgb}{0.284602,0.317763,0.963947}
\newcommand{\vev}[1]{ \left< {#1} \right> }

\newcommand{\dd}{\mathrm{d}}

\newcommand{\ee}{\mathrm{e}}

\makeatletter
\newcommand\footnoteref[1]{\protected@xdef\@thefnmark{\ref{#1}}\@footnotemark}
\makeatother

\begin{document}

%%%%%%%%%%%%%%%%%%%%%%%%%%%
%%%%%%%%%%% Title %%%%%%%%%%%
%%%%%%%%%%%%%%%%%%%%%%%%%%%

%%paper
\title{Circular polarization of the cosmic microwave background
from vector and tensor perturbations}
\author{Keisuke Inomata}
\affiliation{ICRR, University of Tokyo, Kashiwa, 277-8582, Japan}
\affiliation{Kavli IPMU (WPI), UTIAS, University of Tokyo, Kashiwa, 277-8583, Japan}
\author{Marc Kamionkowski}
\affiliation{Department of Physics and Astronomy, Johns Hopkins University, 3400 N. Charles Street,
Baltimore, MD 21218, U.S.A.}

\begin{abstract}
Circular polarization of the cosmic microwave background (CMB) can be
induced by Faraday conversion of the primordial linearly
polarized radiation as it propagates through a birefringent
medium.  Recent work has shown that the dominant source of
birefringence from primordial density perturbations is
the anisotropic background CMB.  Here we extend prior work to allow for the
additional birefringence that may arise from primordial vector
and tensor perturbations.  We derive the formulas for the
power spectrum of the induced
circular polarization and apply those to the standard
cosmology.  We find the root-variance of the induced circular
polarization to be $\sqrt{\vev{V^2}}\sim 3\times 10^{-14}$ for
scalar perturbations and $\sqrt{\vev{V^2}}\sim 7\times
10^{-18} (r/0.06)$ for tensor perturbations with a
tensor-to-scalar ratio $r$.
\end{abstract}

\date{\today}
\maketitle
\preprint{IPMU 18-0184}

%%%%%%%%%%%%%%%%%%%%%%%%%%%%%%
\section{Introduction}
\label{sec:intro}
%%%%%%%%%%%%%%%%%%%%%%%%%%%%%%

The cosmic microwave background (CMB) has helped us understand
the history of the Universe.  Through measurement of the temperature
and polarization fluctuations in the CMB, we have determined
precisely the classical cosmological parameters~\cite{Aghanim:2018eyx}. 
However, the temperature measurements are already limited by cosmic
variance, thus motivating the investigation of other
observables, such as polarization
\cite{Hu:1997hv,Kamionkowski:2015yta,Abazajian:2016yjj} and
frequency distortions \cite{Chluba:2015bqa} of the CMB. 

In this paper, we focus on the circular polarization.
In astrophysics, circular polarization may arise in
masers~\cite{deguchi1986circular,Watson:2001hg}, gamma-ray-burst
afterglows~\cite{Matsumiya:2003pw,Wiersema:2014bha,Batebi:2016efn,Shakeri:2018qal}, jets of active
galactic
nuclei~\cite{Brunthaler:2001dq,Homan:2009wf,Homan:2001yy,Beckert:2001az},
and
pulsars~\cite{Karastergiou:2003sp,melrose2003causes,Mitra:2009mk,Shakeri:2017knk}.
In addition, circular polarization has recently been discussed
also in the context of the CMB~\cite{Bavarsad:2009hm,Batebi:2016ocb,De:2014qza,King:2016exc,Montero-Camacho:2018vgs,Kamionkowski:2018syl}.  Circular polarization can be
produced through Faraday conversion when a linearly polarized
light ray propagates through a medium where the indexes of
refraction differ along the two different transverse axes.
In this way the linear polarization induced at the CMB
last-scattering surface can be converted to circular
polarization.  Refs.~\cite{Bavarsad:2009hm,Batebi:2016ocb} discuss CMB
circular polarization produced by birefringence from
magnetic fields and from new physics beyond the Standard Model (BSM). 
The circular polarization produced via Faraday conversion due to
supernova remnants of Population III stars is discussed in
Ref.~\cite{De:2014qza,King:2016exc}.
The current constraint to the CMB circular-polarization angular
power spectrum is $l(l+1)C_l^{VV}/(2\pi) \lesssim 10^{-8}$ at
multipole moments $l>3000$~\cite{partridge1988linear}, $\lesssim
3\times 10^{-11}$ at $33 < l < 307$~\cite{Nagy:2017csq}, and
$\lesssim10^{-7}$ at larger scale~\cite{Mainini:2013mja}.
Forthcoming experiments, such as
CLASS~\cite{Essinger-Hileman:2014pja} and
PIPER~\cite{Lazear:2014bga}, are expected to improve
considerably on the sensitivity to CMB polarization.  

Recently, a detailed investigation of the circular polarization
that arises from primordial perturbations was presented in
Ref.~\ \cite{Montero-Camacho:2018vgs}.  No circular
polarization arises at linear order, but there are several physical
mechanisms that, at second order in the primordial-perturbation
amplitude, can induce circular polarization from the primordial
linear polarization.  Although this primordially-induced
circular polarization may be smaller than that induced by other
late-time astrophysical effects, and/or BSM physics, these
predictions are more robust and may be thought of as a lower
bound to the expected circular polarization.  There are a number
of possible standard-model sources of the cosmic birefringence
needed for Faraday conversion, including, for example,
spin-polarization of hydrogen atoms induced by an anisotropic
CMB background \cite{Montero-Camacho:2018vgs}.  Still, the most
significant source is
photon-photon interactions~\cite{Montero-Camacho:2018vgs,Motie:2011az, Sawyer:2014maa,
Ejlli:2016avx,Shakeri:2017iph,Sadegh:2017rnr}, which is the mechanism we
consider here.  In this case, the required birefringence is
provided by the CMB anisotropies seen by the CMB photon as it
propagates from the surface of last scatter.

In this paper, we extend prior work by considering the
additional cosmic birefringence that may be induced by
primordial vector and tensor perturbations.  In particular,
tensor perturbations, or primordial gravitational waves,
are a highly sought relic in the canonical single-field
slow-roll inflationary paradigm
\cite{Kamionkowski:2015yta,Abazajian:2016yjj}.  Since the 
tensor contribution to the CMB quadrupole may be almost 10\% of the total,
it is conceivable---given order-unity factors---that the tensor
contribution to the circular polarization may rival the scalar
contribution.  Note that although the photon-graviton scattering can also induce the circular polarization from tensor perturbations~\cite{Bartolo:2018igk},
 the induced circular polarization in CMB is much smaller than that induced through photon-photon scattering as we will see later.
The calculation is also valuable as an
illustrative application of the total-angular-momentum (TAM)
formalism~\cite{Dai:2012bc,Dai:2012ma} employed earlier
\cite{Kamionkowski:2018syl} for the simpler scalar-perturbation
case.  In the TAM formalism, primordial perturbations are
expanded in terms of TAM waves, which are eigenstates of the
generators of rotations, rather than the usual plane waves
(eigenstates of the generators of spatial translations).  The
TAM formalism allows for predictions
for observables on a spherical sky to be obtained far more
simply than through traditional approaches, particularly for
vector and tensor perturbations.

This paper is organized as follows.
In Section~\ref{sec:basic_formulas}, we introduce the basic formulas describing circular polarizations induced through the Faraday conversion.
Then, we briefly review the TAM formalism in Section~\ref{sec:tam_formalism}.
In Section~\ref{sec:cal_phi}, we take the photon-photon scattering source term as a concrete example and show how to express the source term with the TAM formalism.
In Section~\ref{sec:cal_cl_vv}, we relate the source term to the angular power spectrum and perform numerical calculations assuming the standard cosmology.
We make some concluding remarks in Section~\ref{sec:conclusions}.
Note that, throughout this paper, we take the Cartesian coordinate and the metric $g_{ij}$ equals to $\delta_{ij}$.

%%%%%%%%%%%%%%%%%%%%%%%%%%%%%%
\section{Basic formulas for circular polarization}
\label{sec:basic_formulas}
%%%%%%%%%%%%%%%%%%%%%%%%%%%%%%

In this Section, we introduce the formulas for the circular
polarizations induced by Faraday conversion.  Faraday conversion
occurs when a light ray passes through a medium in which each
axis perpendicular to the light-ray trajectory has a  different
index of refraction. The three-dimensional index-of-refraction tensor is given by~\cite{Montero-Camacho:2018vgs}
\begin{align}
n_{ij} = \delta_{ij} + \frac{1}{2} ( \chi_{e,ij} + \chi_{m,ij} ),
\label{eq:n_ij_def}
\end{align}
where $\chi_{e,ij}$ and $\chi_{m,ij}$ are the electric and
magnetic susceptibilities respectively. 
We focus on the $x$ and $y$ components of the tensor ($z$ axis: photon trajectory) because photon does not have the longitudinal polarization.
Then, the index-of-refraction tensor in the two-dimensional plane perpendicular to the trajectory can be expressed with four parameters as
\begin{align} 
n_{ab} &= \left(
    \begin{array}{cc}
     n_I + n_Q & n_U + in_V  \\
     n_U - in_V & n_I - n_Q \\
    \end{array}
  \right).
  \label{eq:n_ab_def}
\end{align}
Here, $n_I$ is the polarization-averaged index of refraction, $n_Q$
the difference between the indexes of refraction in $x$ and
$y$ axes in the transverse plane, and $n_U$ is the difference
between the indexes of refraction on two axes that are
rotated by $45^\circ$ from the $x$ and
$y$ axes.  Also, $n_V$ is
the difference between the indexes of refraction for the two
different circular polarizations, which we ignore in the following because it does not convert linear polarization to circular polarization~\cite{Montero-Camacho:2018vgs}.
The relation between Eqs.~(\ref{eq:n_ij_def}) and (\ref{eq:n_ab_def}) are given as $n_I = \frac{1}{2} (n_{xx} + n_{yy})$, $n_Q = \frac{1}{2} (n_{xx} - n_{yy})$, and $n_U = \frac{1}{2} (n_{xy} + n_{yx})$.
In the following, we use the subscripts $i, j$ and $k$ to describe the three-dimensional space
 and use the subscripts $a, b$ and $c$ to describe the two-dimensional plane perpendicular to the trajectory.

An observed CMB photon has a radial trajectory that
arrives from some observed direction $\hat{\bm n}$.  
According to Refs.~\cite{Montero-Camacho:2018vgs,Kamionkowski:2018syl}, the circular polarization $V(\hat{ \bm{n}})$ observed in
direction with Stokes parameters $Q(\hat{\bm n})$ and $U(\hat{ \bm{n}})$ at
the surface of last scatter is given as
\begin{align}
V(\hat{\bm n}) = \phi_Q(\hat{\bm n}) U(\hat{\bm n}) - \phi_U(\hat{\bm n}) Q(\hat{\bm n}),
\label{eq:v_formula_1}
\end{align}
where the phases $\phi_{Q,U}(\hat{\bm n})$ are obtained as
integrals,
\begin{align}
\phi_{Q,U}(\hat{\bm n}) = \frac{2}{c} \int^{\chi_\text{LSS}}_0 \frac{\dd \chi}{1+z} \, \omega(\chi) n_{Q,U}(\hat{\bm n} \chi, \eta_0 - \chi),
\label{eq:phi_qu_def}
\end{align}
 over comoving distance $\chi$.  Here,
$z$ is redshift, $\chi_\text{LSS}$ is the comoving
distance to the last-scattering surface and $\eta_0$ is the current conformal time.
Note that a general refractive tensor is space-time dependent as $n_{ab}(\bm x, \eta)$.
Although the
linear-polarization pattern on large angular scales is altered
by reionization, the dominant contributions to the phase shift
occur soon after the last scattering (see
Section~\ref{sec:cal_phi}).  Thus, the circular polarization
induced after the reionization is negligible and neglected in
the following.

The Stokes parameters $Q(\hat{\bm n})$ and $U(\hat{\bm n})$, as well as
the phases $\phi_Q(\hat{\bm n})$ and $\phi_U(\hat{\bm n})$, are not
rotational invariants; they are components (in the
$x$-$y$  coordinate system), respectively, of
polarization and phase-shift tensors, which are, respectively,~\cite{Kamionkowski:2018syl}
\begin{equation}
P_{ab}(\hat{\bm n}) = \frac{1}{\sqrt{2}} \left(
    \begin{array}{cc}
     Q(\hat{\bm n}) & U(\hat{\bm n})  \\
     U(\hat{\bm n}) & -Q(\hat{\bm n}) \\
    \end{array}
  \right),
\qquad
\Phi_{ab}(\hat{\bm n}) = \frac{1}{\sqrt{2}} \left(
    \begin{array}{cc}
     \phi_Q(\hat{\bm n}) & \phi_U(\hat{\bm n})  \\
     \phi_U(\hat{\bm n}) & -\phi_Q(\hat{\bm n}) \\
    \end{array}
  \right).
\end{equation}
Then, we can rewrite Eq.~(\ref{eq:v_formula_1}) as
\begin{align}
     V(\hat{\bm n}) = \epsilon_{ac} P^{ab} (\hat{\bm n})
     \Phi_b^{\ c} (\hat{\bm n}),
\label{eq:v_formula_2}
\end{align}
where $\epsilon_{ab}$ is the antisymmetric tensor on the 2-sphere.

%%%%%%%%%%%%%%%%%%%%%%%%%%%%%%
\section{TAM formalism}
\label{sec:tam_formalism}
%%%%%%%%%%%%%%%%%%%%%%%%%%%%%%

In this Section, we briefly review aspects of the
total-angular-momentum (TAM)
formalism~\cite{Dai:2012bc,Dai:2012ma} relevant for this work.
In particular, we focus on the TAM formalism for tensor fields
because the relevant anisotropies in the index of refraction are
described by an index-of-refraction tensor field.  Throughout
this paper, we follow the notation and conventions for the TAM
formalism used in Ref.~\cite{Dai:2012bc}.
In the following, we consider a symmetric trace-free tensor 
because, as we will see in the next Section, the Faraday conversion is only related to the trace-free part of the index-of-refraction tensor.

In the usual approach, a symmetric trace-free tensor field can
be expanded in terms of plane waves of helicities
$\lambda=-2,\ldots,2$ as,
\begin{align}
h_{ij}(\bm x) 
= \sum_{\lambda=-2}^2  \int \frac{\dd^3 k}{(2\pi)^3} h^\lambda
(\bm k) (\hat{\varepsilon}^{\lambda}_{ij}(\hat{\bm k}))^* \ee^{i \bm k \cdot \bm
x},
\label{eq:h_plane_def}
\end{align}
where the power spectra are given by
\begin{align}
\vev{h^\lambda (\bm k) [h^{\lambda'} (\bm k')]^* } = \begin{cases}
      \delta^{\lambda \lambda'} (2\pi)^3 \delta(\bm k - \bm k') P_L(k) & (|\lambda| = 0), \\
      \delta^{\lambda \lambda'} (2\pi)^3 \delta(\bm k - \bm k') P_V(k) & (|\lambda| = 1), \\
      \delta^{\lambda \lambda'} (2\pi)^3 \delta(\bm k - \bm k') P_T(k) & (|\lambda| = 2).
      \label{eq:h_power_spectra}
  \end{cases}
\end{align}
Here, $P_L(k)$, $P_V(k)$, and $P_T(k)$ are the power spectra for
the longitudinal, transverse-vector, and transverse-traceless
components of the tensor field, and $\hat{\varepsilon}^\lambda_{ij}$ are polarization tensors defined as~\cite{Dai:2012bc}
\begin{align}
\hat{\varepsilon}^{\pm 1}_i (\hat {\bm k} ) = \mp \frac{1}{\sqrt{2}} ( \hat{\theta}_i \mp i \hat{\phi}_i ), \quad
\hat{\varepsilon}^{\pm 1}_{ij} (\hat {\bm k} )= \frac{1}{\sqrt{2}} [ \hat{\varepsilon}^{\pm 1}_i \hat{k}_j +  \hat{\varepsilon}^{\pm 1}_j \hat{k}_i ], \quad
\hat{\varepsilon}^{\pm 2}_{ij} (\hat {\bm k} )= - \hat{\varepsilon}^{\pm 1}_i\hat{\varepsilon}^{\pm 1}_j, \quad
\hat{\varepsilon}^{0}_{ij} (\hat {\bm k} )= \sqrt{\frac{3}{2}} \left( \frac{1}{3} \delta_{ij} - \hat{k}_i \hat{k}_j \right),
\label{eq:varep_def}
\end{align}
where $\hat{\bm \theta}$ and $\hat{\bm \phi}$ are the transverse directions of $\hat{\bm k}$.

However, we can alternatively expand a symmetric trace-free
tensor field in terms of total-angular-momentum (TAM) waves
as~\cite{Dai:2012bc},
\begin{align}
   h_{ij}(\bm x) = \sum_{\lambda=0,\pm 1, \pm 2} \sum_{lm} \int \frac{k^2 \dd k}{(2\pi)^3} h^{k,\lambda}_{(lm)} 4\pi i^l \Psi^{k,\lambda}_{(lm)ij}(\bm x) 
=  \sum_{\alpha=L,VE,VB,TE,TB} \sum_{lm} \int \frac{k^2 \dd k}{(2\pi)^3} h^{k,\alpha}_{(lm)} 4\pi i^l \Psi^{k,\alpha}_{(lm)ij}(\bm x).
\label{eq:h_tam_def}
\end{align}
Here we have written the tensor field in terms of longitudinal
(L), vector-E (VE) and vector-B (VB), and tensor-E (TE) and
tensor-B (TB) modes, and then also in terms of an alternative
helicity basis, with $\lambda=-2,\ldots,2$.
The TAM waves are defined as
\begin{align}
     \Psi^{k,L}_{(lm)ij} ({\bm x}) &= \sqrt{\frac{3}{2}} T^L_{ij}
     \Psi^k_{(lm)}({\bm x}), \quad
     \Psi^{k,VE}_{(lm)ij}({\bm x}) = -\sqrt{\frac{2}{l(l+1)}}
     T^{VE}_{ij}\Psi^k_{(lm)}({\bm x}), \quad
     \Psi^{k,VB}_{(lm)ij}({\bm x}) = -\sqrt{\frac{2}{l(l+1)}}
     T^{VB}_{ij}\Psi^k_{(lm)}({\bm x}), \nonumber \\ 
\Psi^{k,TE}_{(lm)ij}({\bm x}) &= - \sqrt{\frac{(l-2)!}{2(l+2)!}}
T^{TE}_{ij}\Psi^k_{(lm)}({\bm x}), \quad \Psi^{k,TB}_{(lm)ij}({\bm
x}) = -\sqrt{\frac{(l-2)!}{2(l+2)!}} T^{TB}_{ij}\Psi^k_{(lm)}({\bm x}),
\end{align}
where $T^{\alpha}_{ij}$ is defined as
\begin{align}
D_i &\equiv  \frac{i}{k} \nabla_i, \quad L_i \equiv -i r
\epsilon_{ijk} \hat{n}^j \nabla^k, \quad K_i \equiv -i L_i,
\quad M_{\perp,i} \equiv  \epsilon_{ijk} D^j K^k, \quad T^L_{ij} \equiv -D_i D_j + \frac{1}{3}
\delta_{ij}, \quad T^{VE}_{ij} \equiv D_{(i}M_{j)}, \nonumber
\\ T^{VB}_{ij} &\equiv D_{(i} K_{j)}, \quad
T^{TE}_{ij} \equiv M_{(i} M_{j)} - K_{(i} K_{j)} + 2D_{(i} M_{j)}, \quad T^{TB}_{ij} \equiv K_{(i} M_{j)} + M_{(i} K_{j)} + 2D_{(i} K_{j)},
\end{align}
and $\Psi^k_{(lm)}({\bm x}) = j_{l}(k\chi) Y_{(lm)}(\hat{\bm n}) \ (\bm x = \chi \bm{\hat n})$
are scalar TAM waves, written in terms of the spherical Bessel
function $j_l(x)$ and spherical harmonic $Y_{(lm)}(\hat{\bm n})$.

The helicity-basis TAM waves are then,
\begin{align}
\Psi^{k,0}_{(lm)ij} ({\bm x}) = \Psi^{k,L}_{(lm) ij}({\bm x}), \quad
\Psi^{k,\pm 1}_{(lm)ij}({\bm x}) = \frac{1}{\sqrt{2}}
\left[\Psi^{k,VE}_{(lm) ij}({\bm x}) \pm i \Psi^{k,VB}_{(lm)
ij}({\bm x}) \right], \quad
\Psi^{k,\pm 2}_{(lm)ij}({\bm x}) = \frac{1}{\sqrt{2}}
\left[\Psi^{k,TE}_{(lm) ij}({\bm x}) \pm i \Psi^{k,TB}_{(lm)
ij}({\bm x}) \right].
\end{align}
The relations between $h^{k,\lambda}_{(lm)}$ and
$h^{k,\alpha}_{(lm)}$ are given by 
\begin{align}
h^{k,0}_{(lm)} &= h^{k,L}_{(lm)}, \quad
h^{k,\pm 1}_{(lm)} = \frac{1}{\sqrt{2}} \left[ h^{k,VE}_{(lm)} \mp  i \, h^{k,VB}_{(lm)} \right], \quad
h^{k,\pm 2}_{(lm)} = \frac{1}{\sqrt{2}} \left[ h^{k,TE}_{(lm)} \mp  i \, h^{k,TB}_{(lm)} \right].
\end{align}
The plane waves with an arbitrary trace-free polarization tensor $\hat \varepsilon_{ij}$, which is a combination of $\hat \varepsilon^{\lambda}_{ij}$ or $\hat \varepsilon^{\alpha}_{ij}$ in general and can be $\hat \varepsilon^{\lambda}_{ij}$ or $\hat \varepsilon^{\alpha}_{ij}$ themselves in some cases, are related to the TAM basis functions as~\cite{Dai:2012bc}
\begin{align}
\label{eq:epsilon_phi_rel}
\hat{\varepsilon}_{ij}(\hat{\bm k}) \ee^{i \bm k \cdot \bm x} = \sum_{\alpha}\sum_{lm} 4\pi i^l B^\alpha_{(lm)} (\hat{\bm k}) \Psi^{k,\alpha}_{(lm)ij}(\bm x)
 = \sum_{\lambda=0, \pm 1, \pm 2}\sum_{lm} 4\pi i^l B^\lambda_{(lm)} (\hat{\bm k}) \Psi^{k,\lambda}_{(lm)ij}(\bm x),
\end{align}
where $\alpha$ runs over $L, VE, VB, TE, TB$, and 
\begin{align}
B^\alpha_{(lm)}(\hat{\bm k}) = \hat{\varepsilon}^{ij}(\hat{\bm k}) Y^{\alpha\ *}_{(lm)ij}(\hat{\bm k}), 
\quad 
B^\lambda_{(lm)}(\hat{\bm k}) = \hat{\varepsilon}^{ij}(\hat{\bm k})
Y^{\lambda\ *}_{(lm)ij}(\hat{\bm k}).
\end{align}
The tensor spherical harmonics $Y^{\alpha}_{(lm)ij}(\hat{\bm
n})$ are defined as\footnote{The two components that live in the
plane of the sky, which we here refer to as TE and TB modes, are
in much of the literature (which considers only these two
components) as E and B.}
\begin{align}
Y^L_{(lm)ij} (\hat{\bm n}) &= \sqrt{\frac{3}{2}} W^L_{ij} Y_{(lm)}(\hat{\bm n}), \nonumber\\
Y^{VE}_{(lm)ij}(\hat{\bm n}) &= -\sqrt{\frac{2}{l(l+1)}} W^{VE}_{ij}Y_{(lm)}(\hat{\bm n}), \quad Y^{VB}_{(lm)ij}(\hat{\bm n}) = -\sqrt{\frac{2}{l(l+1)}} W^{VB}_{ij}Y_{(lm)}(\hat{\bm n}), \nonumber\\
Y^{TE}_{(lm)ij}(\hat{\bm n}) &= -\sqrt{\frac{(l-2)!}{2(l+2)!}} W^{TE}_{ij}Y_{(lm)}(\hat{\bm n}), \quad Y^{TB}_{(lm)ij}(\hat{\bm n}) = -\sqrt{\frac{(l-2)!}{2(l+2)!}} W^{TB}_{ij}Y_{(lm)}(\hat{\bm n}),
\end{align}
where $W^{\alpha}_{ij}$ is defined as
\begin{align}
N_i &\equiv - \hat{n}_i, \quad K_i \equiv -i L_i,  \quad M_{\perp i} \equiv  \epsilon_{ijk} N^j K^k  \quad
W^L_{ij} \equiv -N_iN_j + \frac{1}{3} \delta_{ij}, \quad W^{VE}_{ij} \equiv N_{(i} M_{\perp j)}, \quad W^{VB}_{ij} \equiv N_{(i}K_{j)}, \nonumber \\
W^{TE}_{ij} &\equiv M_{\perp (i} M_{\perp j)} - K_{(i} K_{j)} + 2N_{(i} M_{\perp j)}, \quad W^{TB}_{ij} \equiv K_{(i} M_{\perp j)} + M_{\perp(i} K_{j)} + 2N_{(i} K_{j)}.
\end{align}
In particular, $Y^{TE}_{(lm)ij}$ and $Y^{TB}_{(lm)ij}$ live in the plane perpendicular to $\hat{\bm n}$ and can be expressed as $Y^{TE}_{(lm)ab}$ and $Y^{TB}_{(lm)ab}$ respectively.
The helicity-basis spherical harmonics $Y^{\lambda}_{(lm)ij}(\hat{\bm
n})$ are defined as
\begin{align}
Y^0_{(lm)ij} (\hat{\bm n}) = Y^L_{(lm) ij}(\hat{\bm n}), \quad
Y^{\pm 1}_{(lm)ij}(\hat{\bm n}) = \frac{1}{\sqrt{2}} \left[
Y^{VE}_{(lm) ij}(\hat{\bm n}) \pm i Y^{VB}_{(lm) ij}(\hat{\bm
n}) \right], \quad 
Y^{\pm 2}_{(lm)ij}(\hat{\bm n}) = \frac{1}{\sqrt{2}} \left[
Y^{TE}_{(lm) ij}(\hat{\bm n}) \pm i Y^{TB}_{(lm) ij}(\hat{\bm
n}) \right].
\end{align}
The $Y^\lambda_{(lm)ij}(\hat{\bm k})$ are related to the
spin-weighted spherical harmonics
by~\cite{Newman:1961qr,Goldberg:1966uu} 
\begin{align}
\hat{\varepsilon}^{ij}_\lambda (\hat{\bm k})  \, Y^{\lambda'}_{(lm)ij}  (\hat{\bm k}) = \ _{-\lambda}Y_{(lm)}(\hat{\bm k}) \delta^{\ \lambda'}_\lambda.
\label{eq:varep_y_rel}
\end{align}
From Eqs.~(\ref{eq:h_plane_def}), (\ref{eq:h_tam_def}), and (\ref{eq:varep_y_rel}), we can derive the following relations between $h^\lambda(\bm k) $ and $h^{k,\lambda}_{(lm)}$: 
\begin{align}
h^{k,\lambda}_{(lm)} = \int \dd \hat{\bm k}\, 
h^{\lambda}(\bm k) \left(_{-\lambda}Y_{(lm)}(\hat{\bm k})
\right)^*.
\end{align}
As a result, the TAM amplitudes satisfy\footnote{
The spin-weighted spherical harmonics satisfy  $\int \dd
\hat{\bm n}\  _{\lambda}Y_{(lm)}(\hat{\bm n})
_{\lambda'}Y^*_{(l'm')}(\hat{\bm n}) = \delta_{ll'} \delta_{mm'}
\delta_{\lambda \lambda'}$.}
\begin{align}
  \label{eq:h_lambda_power}
\vev{h^{k,\lambda}_{(lm)}  [h^{k',\lambda'}_{(l'm')} ]^* } &= \begin{cases}
      \delta_{ll'} \delta_{mm'} \delta^{\lambda \lambda'} \frac{(2\pi)^3}{k^2} \delta( k - k') P_L(k) & (|\lambda| = 0), \\
      \delta_{ll'} \delta_{mm'} \delta^{\lambda \lambda'} \frac{(2\pi)^3}{k^2} \delta( k -  k') P_V(k) & (|\lambda| = 1), \\
      \delta_{ll'} \delta_{mm'} \delta^{\lambda \lambda'} \frac{(2\pi)^3}{k^2} \delta( k -  k') P_T(k) & (|\lambda| = 2),     
  \end{cases} \\
  \label{eq:h_alpha_power}
\vev{h^{k,\alpha}_{(lm)} [h^{k',\alpha'}_{(l'm')} ]^* } &= \begin{cases}
      \delta_{ll'} \delta_{mm'} \delta^{\alpha \alpha'} \frac{(2\pi)^3}{k^2} \delta( k - k') P_L(k) & (\alpha = L), \\
      \delta_{ll'} \delta_{mm'} \delta^{\alpha \alpha'} \frac{(2\pi)^3}{k^2} \delta( k -  k') P_V(k) & (\alpha = VE, VB), \\
      \delta_{ll'} \delta_{mm'} \delta^{\alpha \alpha'} \frac{(2\pi)^3}{k^2} \delta( k -  k') P_T(k) & (\alpha = TE, TB).       
  \end{cases}
\end{align}
%%

%%%%%%%%%%%%%%%%%%%%%%%%%%%%%%
\section{Calculation of $\Phi_{ab}(\hat{\bm n})$}
\label{sec:cal_phi}
%%%%%%%%%%%%%%%%%%%%%%%%%%%%%%

Although there are several physical effects that may generate
Faraday conversion, the dominant mechanism, as noted in
Ref.~\cite{Montero-Camacho:2018vgs}, is photon-photon scattering.
The components of the index-of-refraction tensor due to
photon-photon scattering is (see Appendix~\ref{app:p_p_scat}
and Ref.~\cite{Montero-Camacho:2018vgs} for the derivation),
\begin{align}
\label{eq:nq_def}
n_Q(\bm x, \eta) \equiv \frac{1}{2} (n_{xx} (\bm x, \eta) -n_{yy} (\bm x, \eta) ) &\simeq 48 \sqrt{\frac{\pi}{5}} A_e \mu_0 a_{\rm{rad}} T_{\rm{CMB}}^4 \text{Re} \, a_{2,-2}^E (\bm x,\eta),  \\ 
\label{eq:nu_def}
n_U(\bm x, \eta) \equiv \frac{1}{2} (n_{xy} (\bm x, \eta) +n_{yx} (\bm x, \eta) ) &\simeq 48 \sqrt{\frac{\pi}{5}}
A_e \mu_0 a_{\rm{rad}} T_{\rm{CMB}}^4 \text{Im} \, a_{2,-2}^E (\bm x, \eta).
\end{align}
Here, $\mu_0$ is the magnetic permeability of the vacuum,
$a_\text{rad}$ is the radiation energy density constant,
$T_\text{CMB}$ is the CMB temperature, $a^E_{lm}(\bm x, \eta)$ is the
coefficient of the local E-mode moment induced by primordial
perturbations, and $A_e$ is the Euler-Heisenberg interaction
constant, which can be expressed with electron mass $m_e$,
Compton wavelength $\lambda_e$, and the fine structure constant
$\alpha$ as $A_e = 2\alpha^2 \lambda_e^3 /(45\,\mu_0 m_e c^2)$.
Next, we write the spatial dependence of the indexes of
refraction in terms of Fourier components through,
\begin{align}
\label{eq:n_q_f}
n_Q (\bm x, \eta) \simeq 48 \sqrt{\frac{\pi}{5}} A_e \mu_0 a_{\rm{rad}} T_{\rm{CMB}}^4
\frac{1}{2} \int \frac{\dd^3 k}{(2\pi)^3} ( a_{2,-2}^E (\bm k, \eta) +  a_{2,2}^E (\bm k, \eta) ) \ee^{i\bm k \cdot \bm x}, \\
\label{eq:n_u_f}
n_U (\bm x, \eta) \simeq 48 \sqrt{\frac{\pi}{5}} A_e \mu_0 a_{\rm{rad}} T_{\rm{CMB}}^4
\frac{1}{2i} \int \frac{\dd^3 k}{(2\pi)^3} ( a_{2,-2}^E (\bm k, \eta) -  a_{2,2}^E (\bm k, \eta) ) \ee^{i\bm k \cdot \bm x},
\end{align}
where we have used the relation $a^{E *}_{22} = a^{E}_{2,-2}$.
We then transform the local quadrupole moments in
Eqs.~(\ref{eq:n_q_f}) and (\ref{eq:n_u_f}) using
\begin{align}
a^E_{2,\pm 2} (\bm k, \eta) =  \sum_{m=0,\pm 1, \pm 2} D^2_{\pm 2, m} (\pi- \phi_{\bm{k}}, \theta_{\bm{k}}, 0) \bar{a}_{2, m}^E(\bm k, \eta), 
\end{align}
where un-barred quantities are in the line-of-sight frame (with
the z-axis toward the observer), 
and barred quantities are in the wave vector frame ($\bm k$ is
the $z$ axis and the direction toward the observer in the $\bar
x\bar z$-plane, with $\bar x<0$).
Here, $\theta_{\bm k}$ and $\phi_{\bm k}$ are the polar and
azimuthal angle of $\hat {\bm k}$ in the line-of-sight frame,
and $D^l_{m'm}$ is the Wigner rotation matrix, which is defined in
Eq.~(\ref{eq:wigner_d_def}). Then, we derive 
\begin{align}
\label{eq:re_a_2m2}
 a^E_{2,-2} (\bm k, \eta) + a^E_{22} (\bm k, \eta)  =& 
\sum_m D^2_{-2, m} (\pi- \phi_{\bm{k}}, \theta_{\bm{k}}, 0) \bar{a}_{2, m}^E(\bm k, \eta) + \sum_m D^2_{2, m} (\pi- \phi_{\bm{k}}, \theta_{\bm{k}}, 0) \bar{a}_{2, m}^E(\bm k, \eta) \nonumber \\
=& \sqrt{\frac{3}{2}} (\cos \phi^2_{\bm{k}} -\sin \phi^2_{\bm{k}}) \sin \theta^2_{\bm{k}} \bar{a}_{2, 0}^E(\bm k, \eta) -  \sin\theta_{\bm{k}} (-\cos\theta_{\bm{k}} \cos 2\phi_{\bm{k}} + i \sin 2\phi_{\bm{k}})  \bar{a}_{2, 1}^E(\bm k, \eta) \nonumber \\
& -  \sin\theta_{\bm{k}} (\cos\theta_{\bm{k}} \cos 2\phi_{\bm{k}} + i \sin 2\phi_{\bm{k}})  \bar{a}_{2, -1}^E(\bm k, \eta) + \left( \frac{3 + \cos 2\theta_{\bm{k}} }{4} \cos 2\phi_{\bm{k}} - i \cos \theta_{\bm{k}} \sin 2\phi_{\bm{k}} \right) \bar{a}_{2, 2}^E(\bm k, \eta)  \nonumber \\
& + \left( \frac{3 + \cos 2\theta_{\bm{k}} }{4} \cos 2\phi_{\bm{k}} + i \cos \theta_{\bm{k}} \sin 2\phi_{\bm{k}} \right)\bar{a}_{2, -2}^E(\bm k, \eta) \nonumber \\
=& -( (\hat{\varepsilon}^{0}_{xx} (\hat{\bm k}))^* -  (\hat{\varepsilon}^{0}_{yy} (\hat{\bm k}))^* ) \bar{a}_{2, 0}^E(\bm k, \eta)
- ( (\hat{\varepsilon}^{+1}_{xx}  (\hat{\bm k}) )^* -  (\hat{\varepsilon}^{+1}_{yy} (\hat{\bm k}) )^* )  \bar{a}_{2, 1}^E(\bm k, \eta) \nonumber \\
&- ( (\hat{\varepsilon}^{-1}_{xx} (\hat{\bm k}))^*  -  (\hat{\varepsilon}^{-1}_{yy} (\hat{\bm k}))^* ) \bar{a}_{2, -1}^E(\bm k, \eta)
-( (\hat{\varepsilon}^{+2}_{xx} (\hat{\bm k}))^*  - (\hat{\varepsilon}^{+2}_{yy} (\hat{\bm k}))^* ) \bar{a}_{2, 2}^E(\bm k, \eta) \nonumber \\
&- ( (\hat{\varepsilon}^{-2}_{xx} (\hat{\bm k}))^*  - (\hat{\varepsilon}^{-2}_{yy} (\hat{\bm k}))^* ) \bar{a}_{2, -2}^E(\bm k, \eta),
\end{align}
\begin{align}
\label{eq:im_a_2m2}
\frac{1}{i} (a^E_{2,-2} (\bm k, \eta) - a^E_{2, 2} (\bm k, \eta) ) =&
\frac{1}{i} \left(\sum_m D^2_{-2, m} (\pi- \phi_{\bm{k}}, \theta_{\bm{k}}, 0) \bar{a}_{2, m}^E(\bm k, \eta) - \sum_m D^2_{2, m} (\pi- \phi_{\bm{k}}, \theta_{\bm{k}}, 0) \bar{a}_{2, m}^E(\bm k, \eta) \right)\nonumber \\
=&  \sqrt{6} \cos \phi_{\bm{k}}\sin \phi_{\bm{k}} \sin \theta^2_{\bm{k}} \bar{a}_{2, 0}^E(\bm k, \eta) + \sin\theta_{\bm{k}} ( i  \cos 2\phi_{\bm{k}} + \cos \theta \sin 2\phi_{\bm{k}})  \bar{a}_{2, 1}^E(\bm k, \eta) \nonumber \\
& +    \sin\theta_{\bm{k}} ( i  \cos 2\phi_{\bm{k}} - \cos \theta_{\bm k} \sin 2\phi_{\bm{k}})  \bar{a}_{2, -1}^E(\bm k, \eta) \nonumber \\
& +  \left( \frac{3 + \cos 2\theta_{\bm k}}{4} \sin 2\phi_{\bm k} + i \cos \theta_{\bm k} \cos 2\phi_{\bm k} \right) \bar{a}_{2, 2}^E(\bm k, \eta)  \nonumber \\
& 
 +  \left( \frac{3 + \cos 2\theta_{\bm k} }{4} \sin 2\phi_{\bm k} - i \cos \theta_{\bm k} \cos 2\phi_{\bm k} \right) \bar{a}_{2, -2}^E(\bm k, \eta) \nonumber \\
=& -2  (\hat{\varepsilon}^{0}_{xy} (\hat{\bm k}))^*   \bar{a}_{2,
0}^E(\bm k, \eta) - 2  (\hat{\varepsilon}^{+1}_{xy} (\hat{\bm k}))^*
\bar{a}_{2, 1}^E(\bm k, \eta) -  2  (\hat{\varepsilon}^{-1}_{xy}
(\hat{\bm k}) )^* \bar{a}_{2, -1}^E(\bm k, \eta) \nonumber \\
&- 2 (\hat{\varepsilon}^{+2}_{xy} (\hat{\bm k}))^*  \bar{a}_{2, 2}^E(\bm
k)  - 2 (\hat{\varepsilon}^{-2}_{xy} (\hat{\bm k}) )^*  \bar{a}_{2,
-2}^E(\bm k, \eta),
\end{align}
where $\hat{\varepsilon}^{\lambda}_{ij}$ is defined by
Eq.~(\ref{eq:varep_def}), and the basis vectors are
\begin{equation}
\hat{\bm \theta}  (\hat{\bm k} ) = (\cos \theta_{\bm k} \cos
\phi_{\bm k}, \cos \theta_{\bm k} \sin \phi_{\bm k}, -\sin
\theta_{\bm k}), 
\quad \hat{\bm \phi}  (\hat{\bm k} ) = (-\sin
\phi_{\bm k}, \cos \phi_{\bm k}, 0).
\end{equation}
Here, we separate the primordial perturbations and their
transfer functions through $ \bar{a}_{2, \lambda}^E(\bm k, \eta) =
\bar{a}^E_{2,\lambda} (k, \eta) h^{\lambda}(\bm k)$,
where the power spectra of $h^{\lambda}(\bm k)$ are given by
Eq.~(\ref{eq:h_power_spectra}).

From Eqs.~(\ref{eq:nq_def}), (\ref{eq:nu_def}),
(\ref{eq:re_a_2m2}), and (\ref{eq:im_a_2m2}), 
we see that the part of $n_{ij}$ proportional to $(\hat \varepsilon^{\lambda}_{ij})^*$ can describe $n_Q$ and $n_U$, which means that the part related to $n_Q$ and $n_U$ in $n_{ij}$ can be expressed with a trace-free tensor, given as Eq.~(\ref{eq:h_plane_def}).
On the other hand, a nonzero-trace part of $n_{ij}$ is irrelevant to $n_Q$ or $n_U$.
Using the relation between $(\hat \varepsilon^{\lambda}_{ij})^*$ and
$\Psi^{k,\lambda}_{(lm)ij}$ given in
Eq.~(\ref{eq:epsilon_phi_rel}), we can express $n_{ij}$ as
\begin{align}
n_{ij} (\bm x, \eta) =& 48 \sqrt{\frac{\pi}{5}} A_e \mu_0  a_{\rm{rad}} T_{\rm{CMB}}^4
 \int \frac{ k^2 \dd k}{(2\pi)^3} \sum_{lm} \sum_{\lambda=0,\pm1, \pm2} 4\pi i^l \Psi^{k,\lambda}_{(lm)ij}(\bm x) h^{k, \lambda}_{(lm)} (-\bar{a}^E_{2, \lambda} (k, \eta) ) + \tilde n_{ij}(\bm x, \eta) \\
=& 48 \sqrt{\frac{\pi}{5}} A_e \mu_0  a_{\rm{rad}} T_{\rm{CMB}}^4 \int \frac{k^2 \dd k}{(2\pi)^3} \sum_{lm} 4\pi i^l  \left(  
\Psi^{k,L}_{(lm)ij} (\bm x) h^{k, L}_{(lm)} (- \bar{a}^E_{2, 0} (k, \eta) ) \right. \nonumber \\
&+ \Psi^{k,VE}_{(lm)ij} (\bm x) h^{k, VE}_{(lm)}
\left(-\bar{a}^E_{2, 1}(k, \eta) \right)
+  \Psi^{k,VB}_{(lm)ij} (\bm x) h^{k, VB}_{(lm)} \left(-\bar{a}^E_{2, 1}(k, \eta) \right) \nonumber \\
&\left. + \Psi^{k,TE}_{(lm)ij} (\bm x) h^{k, TE}_{(lm)} \left(- \bar{a}^E_{2, 2} (k, \eta) \right)
+  \Psi^{k,TB}_{(lm)ij} (\bm x) h^{k, TB}_{(lm)} \left(
-\bar{a}^E_{2, 2} (k, \eta) \right) \right) + \tilde n_{ij}(\bm x, \eta),
\label{eq:n_ij_decomp}
\end{align}
where $\tilde n_{ij}$ is the part of $n_{ij}$ that is independent of $n_Q$ or $n_U$, and
we have used the relation $\bar{a}^E_{2, \lambda} (k, \eta) =
\bar{a}^E_{2, -\lambda} (k, \eta)$~\cite{Hu:1997hp}. 

To calculate the induced circular polarization, we need to
derive $\Phi_{ab}(\hat{\bm n})$ from $n_{ij}(\bm x)$.  Since
$P_{ab}$ and $\Phi_{ab}$ are $2 \times 2$ symmetric trace-free
tensors in the celestial sphere, we can expand them in terms of
$Y^{TE}_{(lm)ab}(\hat{\bm n})$ and $Y^{TB}_{(lm)ab}(\hat{\bm
n})$ as~\cite{Kamionkowski:1996ks,Zaldarriaga:1996xe},
\begin{align}
\label{eq:p_tensor}
P_{ab}(\hat{\bm n}) &= \sum_{lm} \left[P^E_{lm} Y^{TE}_{(lm)ab}(\hat{\bm n}) + P^B_{lm} Y^{TB}_{(lm)ab}(\hat{\bm n})\right], \\
\label{eq:phi_tensor}
\Phi_{ab}(\hat{\bm n}) &= \sum_{lm} \left[\Phi^E_{lm}
Y^{TE}_{(lm)ab}(\hat{\bm n}) + \Phi^B_{lm}
Y^{TB}_{(lm)ab}(\hat{\bm n}) \right].
\end{align}
To relate $n_{ij}(\bm x)$ to the shift tensor
$\Phi_{ab}(\hat{\bm n})$, we define the projection
operator $\Lambda^{\ \ kl}_{ij}(\hat{\bm n})$ to project $n_{ij}(\bm
x)$ to a spin-2 tensor on the celestial sphere as
\begin{align}
\label{eq:lambda_def}
\Lambda^{\ \ kl}_{ij}(\hat{\bm n}) \equiv P^{\ k}_i P^{\ l}_j - \frac{1}{2} P^{m k} P^{\ l}_{m} \, \delta_{ij},
\end{align}
where $P_{ij}$ is given as $P_{ij} = \delta_{ij} - \hat{n}_i \hat{n}_j$. %and $g_{ij}$ is the metric on the three-dimensional space, corresponding to $\delta_{ij}$ in the Cartesian coordinate.
Then, we can express the shift tensor in terms of
$n_{ij}(\bm x)$ as~\cite{Kamionkowski:2018syl}
\begin{align}
\label{eq:phi_ab_formula}
\Phi_{ij} (\hat{\bm n}) &= \frac{\omega_0}{\sqrt{2}} \frac{2}{c} \int_0^{\chi_{\text{LSS}}} \dd \chi \left[ \Lambda^{\ \ k'l'}_{ij}(\hat{\bm n}) n_{k'l'} ( \hat{\bm n} \chi, \eta_0 - \chi) \right],
\end{align}
where $\omega_0$ is the frequency today.
Note that since $\Phi_{ij}$ lives on the plane perpendicular to $\hat{\bm n}$, we can regard $\Phi_{ij}$ as $\Phi_{ab}$\footnote{
In the line-of-sight frame, $\hat{\bm n}$ is parallel to $z$ axis and $a$ and $b$ run over $x$-$y$ plane in the three-dimensional Cartesian coordinate.
In other words, $\Phi_{ij}$ is non-zero only for $i,j \neq z$ in that frame.
}
and, by definition, $\Lambda^{\ \ k'l'}_{ij} \tilde{n}_{k'l'}=0$ is satisfied.
As we found in Eq.~(\ref{eq:n_ij_decomp}), the spatial
dependence of $n_{ij}(\bm x)$ can be expressed with
$\Psi^{k,\lambda}_{(lm)ij}(\bm x)$  and the projection of
$\Psi^{k,\lambda}_{(lm)ij}(\bm x)$ onto the celestial sphere is
discussed in Refs.~\cite{Kamionkowski:2018syl, Dai:2012bc}.
According to Eq.~(94) in Ref.~\cite{Dai:2012bc}, $\Lambda^{\ \ k'l'}_{ij} \Psi^\lambda_{k'l'} ( \hat{\bm n} \chi)$ are given by
\begin{align}
\Lambda^{\ \ k'l'}_{ij}(\hat{\bm n}) \Psi^{k,L}_{(lm)k'l'}(\hat{\bm n} \chi) &= -\frac{\sqrt{3}}{2} \sqrt{\frac{(l+2)!}{(l-2)!}} \frac{j_l(k \chi) }{(k\chi)^2} Y^{TE}_{(lm)ij} (\hat{\bm n}) \equiv -\sqrt{2} \epsilon^{(0)}_l (k\chi) Y^{TE}_{(lm)ij} (\hat{\bm n}), \\ 
\Lambda^{\ \ k'l'}_{ij}(\hat{\bm n}) \Psi^{k,VE}_{(lm)k'l'}(\hat{\bm n} \chi) &= -\sqrt{(l-1)(l+2)} \left( f_l(k\chi) + 2 \frac{j_l(k\chi)}{(k\chi)^2} \right)Y^{TE}_{(lm)ij} (\hat{\bm n}) \equiv -2  \epsilon^{(1)}_l (k\chi) Y^{TE}_{(lm)ij} (\hat{\bm n}), \\
\Lambda^{\ \ k'l'}_{ij}(\hat{\bm n}) \Psi^{k,TE}_{(lm)k'l'}(\hat{\bm n} \chi) &= -\frac{1}{2} \left( -j_l(k\chi) + g_l(k\chi) + 4f_l(k\chi) + 6 \frac{j_l(k\chi)}{(k\chi)^2} \right) Y^{TE}_{(lm)ij} (\hat{\bm n}) \equiv -2  \epsilon^{(2)}_l (k\chi) Y^{TE}_{(lm)ij} (\hat{\bm n}), \\
\Lambda^{\ \ k'l'}_{ij}(\hat{\bm n}) \Psi^{k,VB}_{(lm)k'l'}(\hat{\bm n} \chi) &= -i \sqrt{(l-1)(l+2)} \frac{j_l(k\chi)}{k\chi} Y^{TB}_{(lm)ij} (\hat{\bm n}) \equiv -2i  \beta^{(1)}_l (k\chi) Y^{TB}_{(lm)ij} (\hat{\bm n}), \\
\Lambda^{\ \ k'l'}_{ij}(\hat{\bm n})
\Psi^{k,TB}_{(lm)k'l'}(\hat{\bm n} \chi) &= -i \left(j_l'(k\chi)
+ 2\frac{j_l(k\chi)}{k\chi} \right)Y^{TB}_{(lm)ij} (\hat{\bm n})
\equiv  -2 i \beta^{(2)}_l (k\chi) Y^{TB}_{(lm)ij} (\hat{\bm
n}), 
\label{eq:phi_project}
\end{align}
where $f_l(x) \equiv
\frac{\dd}{\dd x} \frac{j_l(x)}{x}$, $g_l(x) \equiv -j_l(x) -
2f_l(x) + (l-1)(l+2) \frac{j_l(x)}{x^2}$ and
$\epsilon^{(m)}_l(x)$ and $\beta^{(m)}_l(x)$ are defined as in
Ref.~\cite{Hu:1997hp}. 
Note again that since $Y^{TE}_{(lm)ij}$ and $Y^{TB}_{(lm)ij}$ can be described in the plane perpendicular to $\hat{\bm n}$, we can regard them as $Y^{TE}_{(lm)ab}$ and $Y^{TB}_{(lm)ab}$ respectively.

From Eqs.~(\ref{eq:phi_ab_formula})--(\ref{eq:phi_project}) for
$\Phi^{E/B}_{lm}$ and Ref.~\cite{Tram:2013ima} for
$P^{E/B}_{lm}$, we can rewrite the coefficients in
Eqs.~(\ref{eq:p_tensor}) and (\ref{eq:phi_tensor})
as\footnote{Note that Eq.~(\protect\ref{eq:phi_l_exp}) corrects Eq.~(30)
in Ref.~\protect\cite{Kamionkowski:2018syl}.}
\begin{align}
P^{E/B}_{lm} &= \sum_\alpha P^{E/B,\lambda}_{lm}, \quad \Phi^{E/B}_{lm} = \sum_\alpha \Phi^{E/B,\lambda}_{lm}, \\
\label{eq:p_l_exp}
P^{E,L}_{lm} &= 4\pi \int \frac{k^2 \dd k}{(2\pi)^3}
\int^{\eta_0}_0 \dd \eta \, g(\eta) \left(-\sqrt{6} {\cal P}^{(0)} (k, \eta) \right) h^{k,(L)}_{lm} \epsilon^{(0)}_l (k(\eta_0-\eta)), \\
P^{E/B,VE/VB}_{lm} &= 4\pi \int \frac{k^2 \dd k}{(2\pi)^3}
\int^{\eta_0}_0 \dd \eta \, g(\eta) \sqrt{2} \left(-\sqrt{6}
{\cal P}^{(1)} (k, \eta) \right) h^{k,(VE/VB)}_{lm} \phi^{(1)}_l (k(\eta_0-\eta)), \\
P^{E/B,TE/TB}_{lm} &= 4\pi \int \frac{k^2 \dd k}{(2\pi)^3}
\int^{\eta_0}_0 \dd \eta \, g(\eta) \sqrt{2} \left(-\sqrt{6}
{\cal P}^{(2)} (k, \eta) \right) h^{k,(TE/TB)}_{lm} \phi^{(2)}_l (k(\eta_0-\eta)), \\
\label{eq:phi_l_exp}
\Phi^{E,L}_{lm} &= 4\pi A \int \frac{k^2 \dd k}{(2\pi)^3}  \int^{\eta_0}_{\eta_\text{LSS}} \dd \eta \,(1+z)^4 \left(  \bar{a}^E_{2, 0} (k, \eta)  \right) h^{k,(L)}_{lm} \epsilon^{(0)}_l (k(\eta_0-\eta)), \\
\Phi^{E/B,VE/VB}_{lm} &= 4\pi A\int \frac{k^2 \dd k}{(2\pi)^3}  \int^{\eta_0}_{\eta_\text{LSS}} \dd \eta\, (1+z)^4 \sqrt{2} \left(  \bar{a}^E_{2, 1} (k, \eta)  \right) h^{k,(VE/VB)}_{lm} \phi^{(1)}_l (k(\eta_0-\eta)), \\
\label{eq:phi_t_exp}
\Phi^{E/B,TE/TB}_{lm} &= 4\pi A\int \frac{k^2 \dd k}{(2\pi)^3}  \int^{\eta_0}_{\eta_\text{LSS}} \dd \eta\, (1+z)^4 \sqrt{2} \left(  \bar{a}^E_{2, 2} (k, \eta)  \right) h^{k,(TE/TB)}_{lm} \phi^{(2)}_l (k(\eta_0-\eta)),
\end{align}
where the integration variable is changed as $\chi \rightarrow \eta$ \,$(\eta = \eta_0 - \chi)$, $\phi^{m}(x)$ is $\epsilon^{m}(x)$ for $VE$ and $TE$ or $i \beta^{m}(x)$ for $VB$ and $TB$, $g(\eta)$ is the visibility function,
$\mathcal P^{(m)}(k,\eta)$ is the function defined in
Ref.~\cite{Tram:2013ima}, and $A
\equiv 96 (\pi/5)^{1/2} A_e \mu_0  a_{\rm{rad}} T_{\rm{0}}^4
c^{-1} \omega_0 = 1.11 \times 10^{-38} \left( \nu_0/100\,
\text{GHz} \right)\, \text{m}^{-1}$~\cite{Montero-Camacho:2018vgs} .
The power spectra of $h^{k, \alpha}_{lm}$ are given by
Eq.~(\ref{eq:h_alpha_power}). The factor of $(1+z)^4$ implies
that the dominant contribution to the phase shift is near the
epoch of recombination, as we mentioned in
Section~\ref{sec:basic_formulas}.

Finally, we summarize the results for the angular power spectra.
We can express $C_l^{P^E P^E}$ and $C_l^{P^B P^B}$ as
\begin{align}
C_l^{P^E P^E} =& \sum_{\alpha=L,VE,TE} \vev{P^{E,(\alpha)
*}_{lm} P^{E,(\alpha)}_{lm}} = 4\pi \int \frac{\dd k}{k}
  \left(\frac{k^3}{2\pi^2}P^{(L)} (k) \right) \left| \int^{\eta_0}_0 \dd \eta \, g(\eta)
\left(-\sqrt{6} {\cal P}^{(0)} (k, \eta) \right) \epsilon^{(0)}_l (k(\eta_0-\eta)) \right|^2 \nonumber \\
 & + 4\pi \int \frac{\dd k}{k} 2 \left(\frac{k^3}{2\pi^2}P^{(VE)} (k) \right) \left|
 \int^{\eta_0}_0 \dd \eta \, g(\eta) \left(-\sqrt{6} {\cal P}^{(1)} (k, \eta) \right) \epsilon^{(1)}_l (k(\eta_0-\eta) ) \right|^2 \nonumber \\
  & + 4\pi \int \frac{\dd k}{k} 2  \left(\frac{k^3}{2\pi^2}P^{(TE)} (k) \right) \left|
  \int^{\eta_0}_0 \dd \eta \, g(\eta) \left(-\sqrt{6} {\cal
  P}^{(2)} (k, \eta) \right) \epsilon^{(2)}_l (k(\eta_0-\eta) )
  \right|^2,
\end{align}
\begin{align}
  C_l^{P^B P^B} =& \sum_{\alpha=VB,TB} \vev{P^{B,(\alpha) *}_{lm} P^{B,(\alpha)}_{lm}}   
 = 4\pi \int \frac{\dd k}{k} 2 \left(\frac{k^3}{2\pi^2}P^{(VB)} (k) \right) \left|
 \int^{\eta_0}_0 \dd \eta \, g(\eta) \left(-\sqrt{6} {\cal P}^{(1)} (k, \eta) \right) \beta^{(1)}_l (k(\eta_0-\eta) ) \right|^2 \nonumber \\
  & + 4\pi \int \frac{\dd k}{k} 2  \left(\frac{k^3}{2\pi^2}P^{(TB)} (k) \right) \left|
  \int^{\eta_0}_0 \dd \eta \, g(\eta) \left(-\sqrt{6} {\cal P}^{(2)} (k, \eta) \right) \beta^{(2)}_l (k(\eta_0-\eta) ) \right|^2,
\end{align}
where the tensor-to-scalar ratio is defined as $r  = 2 P^{(TE)}(k)/ P^{(L)} (k) = 2 P^{(TB)}(k)/P^{(L)} (k)$. 
Similarly, $C_l^{\Phi^E \Phi^E}$ and $C_l^{\Phi^B \Phi^B}$ are given by 
\begin{align}
C_l^{\Phi^E \Phi^E} =& \sum_{\alpha=L,VE,TE} \vev{\Phi^{E,(\alpha) *}_{lm} \Phi^{E,(\alpha)}_{lm}} = 4 \pi A^2 \left( \int \frac{\dd k}{k}  \left(\frac{k^3}{2\pi^2}P^{(L)} (k) \right) \left| \int^{\eta_0}_{\eta_{\text{LSS}}} \dd \eta \,  (1+z)^4 \left( \bar{a}^E_{2, 0} (k, \eta) \right) \epsilon^{(0)}_l (k(\eta_0-\eta) ) \right|^2 \right. \nonumber\\
&\qquad \ +\int \frac{\dd k}{k} 2 \left(\frac{k^3}{2\pi^2}P^{(VE)} (k) \right) \left| \int^{\eta_0}_{\eta_{\text{LSS}}} \dd \eta \,  (1+z)^4 \left( \bar{a}^E_{2, 1} (k, \eta) \right) \epsilon^{(1)}_l (k(\eta_0-\eta) ) \right|^2 \nonumber \\
&\qquad \ \left. +\int \frac{\dd k}{k} 2 \left(\frac{k^3}{2\pi^2}P^{(TE)} (k) \right) \left|
\int^{\eta_0}_{\eta_{\text{LSS}}} \dd \eta \, (1+z)^4 \left(
\bar{a}^E_{2, 2} (k, \eta) \right) \epsilon^{(2)}_l
(k(\eta_0-\eta) ) \right|^2 \right),
\end{align}
\begin{align}
C_l^{\Phi^B \Phi^B} =& \sum_{\alpha=VB, TB} \vev{\Phi^{B,(\alpha) *}_{lm} \Phi^{B,(\alpha)}_{lm}} = 4 \pi  A^2 \left( \int \frac{\dd k}{k} 2  \left(\frac{k^3}{2\pi^2}P^{(VB)} (k) \right) \left| \int^{\eta_0}_{\eta_{\text{LSS}}} \dd \eta \,  (1+z)^4 \left( \bar{a}^E_{2, 1} (k, \eta) \right) \beta^{(1)}_l (k(\eta_0-\eta) ) \right|^2 \right. \nonumber \\
&\qquad \ \left. +\int \frac{\dd k}{k} 2 \left(\frac{k^3}{2\pi^2}P^{(TB)} (k) \right) \left| \int^{\eta_0}_{\eta_{\text{LSS}}} \dd \eta \, (1+z)^4  \left( \bar{a}^E_{2, 2} (k, \eta) \right) \beta^{(2)}_l (k(\eta_0-\eta) ) \right|^2 \right).
\end{align}
The cross correlations $C_l^{P^E \Phi^E}$ and $C_l^{P^B \Phi^B}$ are also given by
\begin{align}
C_l^{P^E \Phi^E} =& \sum_{\alpha=L,VE,TE} \vev{P^{E,(\alpha) *}_{lm} \Phi^{E,(\alpha)}_{lm}} = 4 \pi A \left( \int \frac{\dd k}{k}  \left(\frac{k^3}{2\pi^2}P^{(L)} (k) \right) \left( \int^{\eta_0}_{\eta_{\text{LSS}}} \dd \eta \, (1+z)^4  \left( \bar{a}^E_{2, 0} (k, \eta) \right) \epsilon^{(0)}_l (k(\eta_0-\eta) ) \right) \right. \nonumber \\
& \qquad\qquad\qquad\qquad\qquad\qquad\qquad\qquad
\times \left(  \int^{\eta_0}_0 \dd \eta \, g(\eta)
\left(-\sqrt{6} {\cal P}^{(0)} (k, \eta) \right) \epsilon^{(0)}_l (k(\eta_0-\eta)) \right) \nonumber\\
&\quad \ +\int \frac{\dd k}{k} 2  \left(\frac{k^3}{2\pi^2}P^{(VE)} (k) \right)  \left( \int^{\eta_0}_{\eta_{\text{LSS}}} \dd \eta \, (1+z)^4 \left( \bar{a}^E_{2, 1} (k, \eta) \right) \epsilon^{(1)}_l (k(\eta_0-\eta) ) \right) \nonumber \\
& \qquad\qquad\qquad\qquad\qquad\qquad\qquad\qquad
 \times \left(  \int^{\eta_0}_0 \dd \eta \, g(\eta)
 \left(-\sqrt{6} {\cal P}^{(1)} (k, \eta) \right) \epsilon^{(1)}_l (k(\eta_0-\eta)) \right) \nonumber \\
&\quad \ +\int \frac{\dd k}{k} 2 \left(\frac{k^3}{2\pi^2}P^{(TE)} (k) \right)  \left( \int^{\eta_0}_{\eta_{\text{LSS}}} \dd \eta \,(1+z)^4 \left( \bar{a}^E_{2, 2} (k, \eta) \right) \epsilon^{(2)}_l (k(\eta_0-\eta) ) \right) \nonumber \\
& \qquad\qquad\qquad\qquad\qquad\qquad\qquad\qquad
\left. \times \left(  \int^{\eta_0}_0 \dd \eta \, g(\eta)
\left(-\sqrt{6} {\cal P}^{(2)} (k, \eta) \right) \epsilon^{(2)}_l
(k(\eta_0-\eta)) \right) \right),
\end{align}
\begin{align}
C_l^{P^B \Phi^B} =& \sum_{\alpha=VB, TB} \vev{P^{B,(\alpha) *}_{lm} \Phi^{B,(\alpha)}_{lm}}  = 4 \pi A \left( \int \frac{\dd k}{k} 2 \left(\frac{k^3}{2\pi^2}P^{(VB)} (k) \right)  \left( \int^{\eta_0}_{\eta_{\text{LSS}}} \dd \eta \,(1+z)^4  \left( \bar{a}^E_{2, 1} (k, \eta) \right) \beta^{(1)}_l (k(\eta_0-\eta) ) \right) \right. \nonumber \\
& \qquad\qquad\qquad\qquad\qquad\qquad\qquad\qquad
  \times \left(  \int^{\eta_0}_0 \dd \eta \, g(\eta)
  \left(-\sqrt{6} {\cal P}^{(1)} (k, \eta) \right) \beta^{(1)}_l (k(\eta_0-\eta)) \right)  \nonumber \\
&\quad \  +\int \frac{\dd k}{k} 2  \left(\frac{k^3}{2\pi^2}P^{(TB)} (k) \right)  \left( \int^{\eta_0}_{\eta_{\text{LSS}}} \dd \eta \, (1+z)^4 \left( \bar{a}^E_{2, 2} (k, \eta) \right) \beta^{(2)}_l (k(\eta_0-\eta) ) \right) \nonumber \\
& \qquad\qquad\qquad\qquad\qquad\qquad\qquad\qquad
\left. \times \left(  \int^{\eta_0}_0 \dd \eta \, g(\eta)
\left(-\sqrt{6} {\cal P}^{(2)} (k, \eta) \right) \beta^{(2)}_l (k(\eta_0-\eta)) \right) \right).
\end{align}
%%

%%%%%%%%%%%%%%%%%%%%%%%%%%%%%%
\section{Calculation of $C_l^{VV}$ and numerical results}
\label{sec:cal_cl_vv}
%%%%%%%%%%%%%%%%%%%%%%%%%%%%%%

In this Section, we explain how to calculate the power spectrum $C_l^{VV}$
for the induced circular polarization with the results
derived in the previous Section, and we present numerical
results for a scale-invariant spectrum of primordial tensor
perturbations.  We follow the discussion in Ref.~\cite{Kamionkowski:2018syl},
but take into account the B mode, which is
not considered in Ref.~\cite{Kamionkowski:2018syl}.
This is because, unlike scalar perturbations, vector and tensor
perturbations induce B modes as we saw in the previous Section.

The angular power spectrum is defined by $C_l^{VV} =
\vev{V^*_{lm} V_{lm}}$ in terms of expansion coefficients,
\begin{equation}
     V_{lm} = \int \dd \hat{\bm n}\,
V(\hat{\bm n}) Y^*_{lm}(\hat{\bm n}).
\label{eq:a_lm_v}
\end{equation}
Substituting Eqs.~(\ref{eq:v_formula_2}), (\ref{eq:p_tensor}),
and (\ref{eq:phi_tensor}) into Eq.~(\ref{eq:a_lm_v}), we obtain
\begin{align}
V_{lm} = \sum_{l_1 m_1} \sum_{l_2 m_2} 
( &P_{l_1 m_1}^E \Phi_{l_2 m_2}^E \int \dd \hat{\bm n}\,
\epsilon^{ab} Y^E_{(l_1 m_1)ac} (\hat{\bm n}) Y^E_{(l_2
m_2)b}\,^c(\hat{\bm n}) Y^*_{lm}(\hat{\bm n}) + P_{l_1 m_1}^E \Phi_{l_2 m_2}^B \int \dd \hat{\bm n}\, \epsilon^{ab} Y^E_{(l_1 m_1)ac} (\hat{\bm n}) Y^B_{(l_2 m_2)b}\,^c(\hat{\bm n}) Y^*_{lm}(\hat{\bm n}) \nonumber \\
+&P_{l_1 m_1}^B \Phi_{l_2 m_2}^E \int \dd \hat{\bm n}\, \epsilon^{ab} Y^B_{(l_1 m_1)ac} (\hat{\bm n}) Y^E_{(l_2 m_2)b}\,^c(\hat{\bm n}) Y^*_{lm}(\hat{\bm n}) \nonumber 
+P_{l_1 m_1}^B \Phi_{l_2 m_2}^B \int \dd \hat{\bm n}\, \epsilon^{ab} Y^B_{(l_1 m_1)ac} (\hat{\bm n}) Y^B_{(l_2 m_2)b}\,^c(\hat{\bm n}) Y^*_{lm}(\hat{\bm n}) ) \\
\label{eq:v_formula2}
= \sum_{l_1 m_1} \sum_{l_2 m_2} 
( &P_{l_1 m_1}^E \Phi_{l_2 m_2}^E \int \dd \hat{\bm n}\, Y^B_{(l_1
m_1)bc} (\hat{\bm n}) Y^E_{(l_2 m_2)}\,^{bc}(\hat{\bm n})
Y^*_{lm}(\hat{\bm n}) + P_{l_1 m_1}^E \Phi_{l_2 m_2}^B \int \dd \hat{\bm n}\, Y^B_{(l_1 m_1)bc} (\hat{\bm n}) Y^B_{(l_2 m_2)}\,^{bc}(\hat{\bm n}) Y^*_{lm}(\hat{\bm n}) \nonumber \\
+&P_{l_1 m_1}^B \Phi_{l_2 m_2}^E \int \dd \hat{\bm n}\,  (-1)
Y^B_{(l_1 m_1)bc} (\hat{\bm n}) Y^B_{(l_2 m_2)}\,^{bc}(\hat{\bm
n}) Y^*_{lm}(\hat{\bm n})+ P_{l_1 m_1}^B \Phi_{l_2 m_2}^B \int \dd
\hat{\bm n}\,   Y^B_{(l_1 m_1)bc} (\hat{\bm n}) Y^E_{(l_2
m_2)}\,^{bc}(\hat{\bm n}) Y^*_{lm}(\hat{\bm n}) ),
\end{align}
where we have used \cite{Gluscevic:2009mm} $\epsilon^{ab}
Y^E_{(l m)ac} (\hat{\bm n}) = Y^B_{(l m)}\,^b \,_c (\hat{\bm
n})$ and $\epsilon^{ab} Y^B_{(l m)ac} (\hat{\bm n}) 
= -Y^E_{(l m)}\,^b \,_c (\hat{\bm n})$.
According to Refs.~\cite{Gluscevic:2009mm, Hu:2000ee}, we can
express the integrals as 
\begin{align}
\label{eq:ybsyb_formula2}
 \int \dd \hat{\bm n}\, Y^{B}_{(l_1 m_1)}\,^{ab} (\hat{\bm n}) Y^B_{(l_2 m_2)ab} (\hat{\bm n})  Y^*_{lm}(\hat{\bm n}) 
 &= \int \dd \hat{\bm n}\, Y^{B*}_{(l_1 \,-m_1)}\,^{ab} (\hat{\bm
 n}) Y^B_{(l_2 m_2)ab} (\hat{\bm n})  Y_{l\,-m}(\hat{\bm n})  =
 \xi^{l\,-m}_{l_1\, -m_1 l_2 m_2} H^l_{l_1 l_2}, \\ 
 \label{eq:yesyb_formula2}
 \int \dd \hat{\bm n}\, Y^{E}_{(l_1 m_1)}\,^{ab} (\hat{\bm n})Y^B_{(l_2 m_2)ab} (\hat{\bm n})  Y^*_{lm}(\hat{\bm n})
  &=  \int \dd \hat{\bm n}\, Y^{E*}_{(l_1 \, -m_1)}\,^{ab}
  (\hat{\bm n}) Y^B_{(l_2 m_2)ab} (\hat{\bm n}) Y_{l\,
  -m}(\hat{\bm n}) = i \xi^{l\, -m}_{l_1\,-m_1 l_2m_2} H^l_{l_1
  l_2},
\end{align}
where the result is zero unless $l_1 + l_2 + l =
(\rm{even})$ in Eq.~(\ref{eq:ybsyb_formula2}) or $l_1 + l_2 + l
= (\rm{odd})$ in Eq.~(\ref{eq:yesyb_formula2}), and $
\xi^{lm}_{l_1m_1 l_2m_2}$ and $ H^l_{l_1 l_2}$ are defined in
terms of Wigner 3-j symbols as
\begin{align}
 \xi^{lm}_{l_1m_1 l_2m_2} \equiv (-1)^m \sqrt{ \frac{(2l_1+1) ( 2l +1) (2l_2 +1)}{4\pi}} 
 \left( 
 \begin{array}{ccc}
      l_1 & l & l_2  \\
      -m_1 & m & m_2
    \end{array}
  \right) , \quad
  H^l_{l_1 l_2} \equiv  \left( 
 \begin{array}{ccc}
      l_1 & l & l_2  \\
      2 & 0 & -2
    \end{array}
  \right).
\end{align}
Here, we define $G^{lm}_{l_1 m_1 l_2 m_2} = - \xi^{lm}_{l_1\, -m_1,
l_2 m_2} H^l_{l_1 l_2}$ (in agreement with the conventions of
Ref.~\cite{Kamionkowski:2018syl}).
Then, using $ \int \dd \hat{\bm n}\, Y^{B}_{(l_1
m_1)}\,^{ab} Y^E_{(l_2 m_2)ab} (\hat{\bm n})  Y^*_{lm}(\hat{\bm
n})= - \int \dd \hat{\bm n}\, Y^{E}_{(l_1 m_1)}\,^{ab} Y^B_{(l_2
m_2)ab} (\hat{\bm n}) Y^*_{lm}(\hat{\bm n})$, we rewrite
Eq.~(\ref{eq:v_formula2}) as
\begin{align}
V_{lm} =& \sum_{l_1 m_1 l_2 m_2 (\rm{odd})}
\left( P_{l_1 m_1}^E \Phi_{l_2 m_2}^E ( i G^{l\, -m}_{l_1 m_1 l_2 m_2} ) + P_{l_1 m_1}^B \Phi_{l_2 m_2}^B ( i G^{l\, -m}_{l_1 m_1 l_2 m_2} ) \right) \nonumber \\
&+  \sum_{l_1 m_1 l_2 m_2 (\rm{even})} 
\left( P_{l_1 m_1}^E \Phi_{l_2 m_2}^B ( -G^{l\, -m}_{l_1 m_1 l_2
m_2} )  - P_{l_1 m_1}^B \Phi_{l_2 m_2}^E ( -G^{l\, -m}_{l_1 m_1
l_2 m_2} ) \right),
\label{eqn:vlm}
\end{align}
where the subscript $(\rm{odd})$ and $(\rm{even})$ means that
the summation is over $l_1 + l_2 + l = (\rm{odd})$
and $l_1 + l_2 + l = (\rm{even})$, respectively. 
Then we derive 
\begin{align}
C_l^{VV} =& \sum_{l_1 m_1 l_2 m_2 (\rm{odd})}
\left[ \left( C_{l_1}^{P^E P^E} C_{l_2}^{\Phi^E \Phi^E} -
C_{l_1}^{P^E \Phi^E} C_{l_2}^{P^E \Phi^E}   \right)  + \left(
C_{l_1}^{P^B P^B} C_{l_2}^{\Phi^B \Phi^B} - C_{l_1}^{P^B \Phi^B}
C_{l_2}^{P^B \Phi^B}   \right) \right] |G^{l\, -m}_{l_1 m_1 l_2
m_2}|^2\nonumber \\
&+  \sum_{l_1 m_1 l_2 m_2 (\rm{even})} 
\left( C_{l_1}^{P^E P^E} C_{l_2}^{\Phi^B \Phi^B} + C_{l_1}^{P^B P^B} C_{l_2}^{\Phi^E \Phi^E} - 2 C_{l_1}^{P^E \Phi^E} C_{l_2}^{P^B \Phi^B} \right) |G^{l\, -m}_{l_1 m_1 l_2 m_2}|^2\\
=& \sum_{l_1 l_2 (\rm{odd})} 
\frac{(2l_1 + 1 )(2l_2 + 1)}{4 \pi} \left( \left( C_{l_1}^{P^E P^E} C_{l_2}^{\Phi^E \Phi^E} - C_{l_1}^{P^E \Phi^E} C_{l_2}^{P^E \Phi^E}   \right) \right. \nonumber \\
 & \qquad \qquad  \qquad \qquad \qquad \qquad \qquad \left. + \left( C_{l_1}^{P^B P^B} C_{l_2}^{\Phi^B \Phi^B} - C_{l_1}^{P^B \Phi^B} C_{l_2}^{P^B \Phi^B}   \right) \right) |H^l_{l_1 l_2}|^2\nonumber \\
&+  \sum_{l_1 l_2 (\rm{even})} 
\frac{(2l_1 + 1 )(2l_2 + 1)}{4 \pi} \left( C_{l_1}^{P^E P^E} C_{l_2}^{\Phi^B \Phi^B} + C_{l_1}^{P^B P^B} C_{l_2}^{\Phi^E \Phi^E} - 2 C_{l_1}^{P^E \Phi^E} C_{l_2}^{P^B \Phi^B} \right) |H^l_{l_1 l_2}|^2\\
\simeq&  \int \frac{\dd^2 l_1}{(2\pi)^2} \sin^2 2 \varphi_{ \bm{l}_1, \bm l - \bm{l}_1}
 \left( \left( C_{l_1}^{P^E P^E} C_{|\bm{l} - \bm{l}_1|}^{\Phi^E
 \Phi^E} - C_{l_1}^{P^E \Phi^E} C_{|\bm{l} - \bm{l}_1|}^{P^E
 \Phi^E}   \right) + \left( C_{l_1}^{P^B P^B} C_{|\bm{l} -
 \bm{l}_1|}^{\Phi^B \Phi^B} - C_{l_1}^{P^B \Phi^B} C_{|\bm{l} -
 \bm{l}_1|}^{P^B \Phi^B}   \right) \right)\nonumber \\
&+  \int \frac{\dd^2 l_1}{(2\pi)^2} \cos^2 2 \varphi_{ \bm{l}_1,\bm{l} - \bm{l}_1}
\left( C_{l_1}^{P^E P^E} C_{|\bm{l} - \bm{l}_1|}^{\Phi^B \Phi^B} + C_{l_1}^{P^B P^B} C_{|\bm{l} - \bm{l}_1|}^{\Phi^E \Phi^E} - 2 C_{l_1}^{P^E \Phi^E} C_{|\bm{l} - \bm{l}_1|}^{P^B \Phi^B} \right),
\label{eq:c_l_vv_final}
\end{align}
where we have used $\sum_{m_1m_2}( \xi^{lm}_{l_1 m_1 l_2 m_2}
)^2 = (2l+1)(2l'+1)/(4\pi)$ \cite{Gluscevic:2009mm, Hu:2000ee}
between the first equality and second equality
and we have used \cite{Gluscevic:2009mm}
\begin{align}
\sum_{l_1 l_2(\text{odd})} \frac{(2l+1)(2l'+1)}{4\pi} |H^l_{l_1 l_2}|^2 &\simeq \int \frac{\dd^2 l_1}{(2\pi)^2} \int \frac{\dd^2 l_2}{(2\pi)^2}  (2\pi)^2 \sin^2 2 \varphi_{ \bm{l}_1,\bm{l}_2} \,\delta(\bm{l} - (\bm{l}_1 + \bm{l}_2) ),\\
\sum_{l_1 l_2(\text{even})} \frac{(2l+1)(2l'+1)}{4\pi} |H^l_{l_1 l_2}|^2 &\simeq \int \frac{\dd^2 l_1}{(2\pi)^2} \int \frac{\dd^2 l_2}{(2\pi)^2} (2\pi)^2 \cos^2 2 \varphi_{ \bm{l}_1,\bm{l}_2} \,\delta(\bm{l} - (\bm{l}_1 + \bm{l}_2) ),
\end{align}
valid when $l,l_1,l_2 \gg 1$, between the second and third equality.

The variance of $V$ is given by $\vev{V^2} = \sum_l (2l+1)
C_l^{VV}/(4\pi)$, which can be approximated,
\begin{align}
\vev{V^2} \simeq  \int \frac{\dd^2 l}{(2\pi)^2} C_l^{VV} \simeq  \frac{1}{2} (\vev{P^E P^E} \vev{\Phi^E \Phi^E} - \vev{P^E \Phi^E}^2 + \vev{P^B P^B} \vev{\Phi^B \Phi^B} - \vev{P^B \Phi^B}^2 \nonumber \\
+ \vev{P^E P^E}\vev{\Phi^B \Phi^B} + \vev{P^B P^B}\vev{\Phi^E \Phi^E} - 2 \vev{P^E \Phi^E}\vev{P^B \Phi^B} ).
\label{eq:vv_vari}
\end{align}

\begin{figure}
\centering
\includegraphics[width=0.7\textwidth]{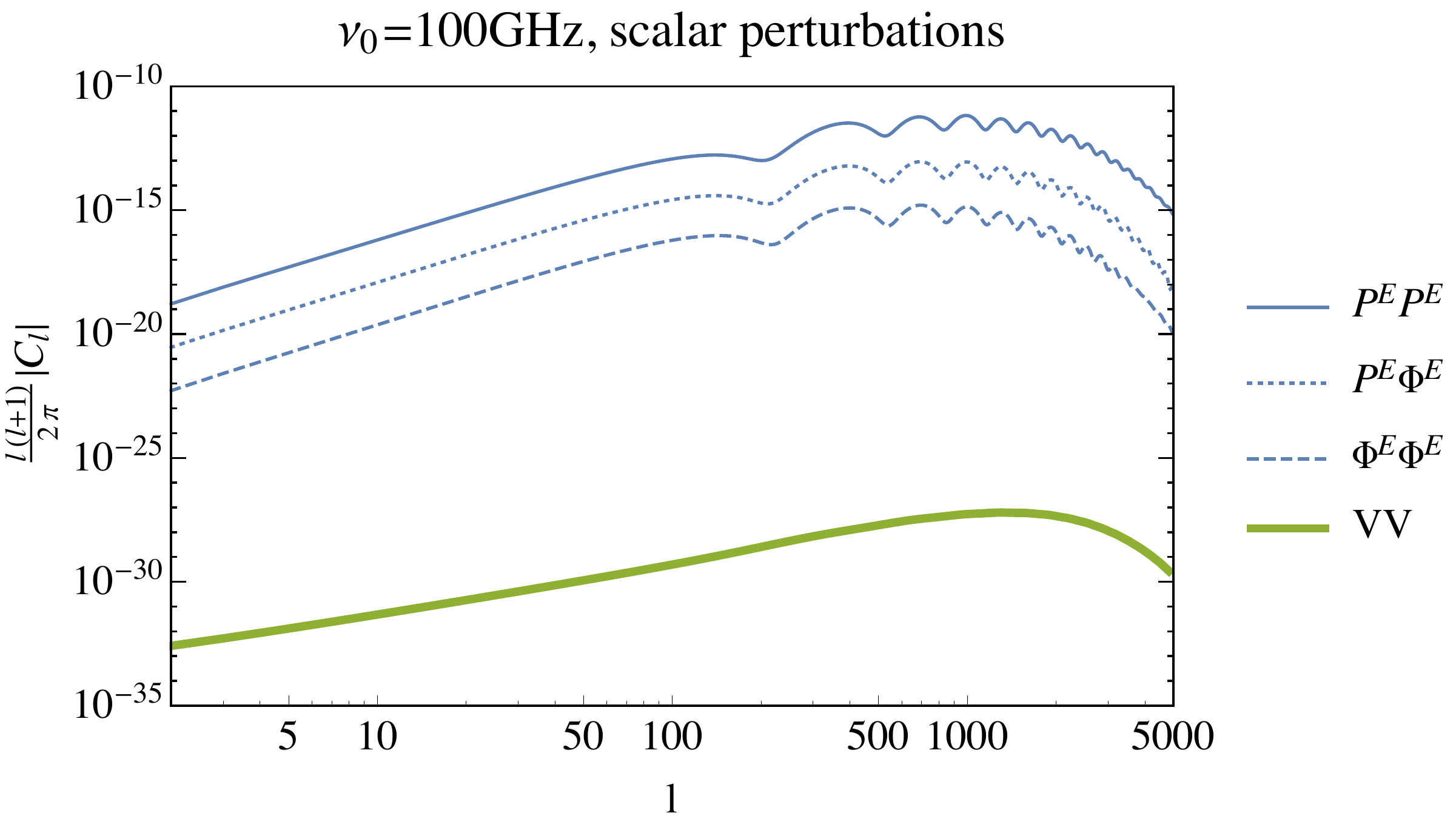}
\caption{The power spectra $C_l$ for scalar perturbations. 
     The blue solid curve shows $C_l^{P^EP^E}$; the blue dotted
     curve shows $|C_l^{P^E \Phi^E}|$; the blue dashed curve shows
     $C_l^{\Phi^E \Phi^E}$; and the green thick solid curve shows
     $C_l^{VV}$. We take $\nu_0=100$~GHz.}
\label{fig:cl_vv_s}
\end{figure}

\begin{figure}
\centering
\includegraphics[width=0.7\textwidth]{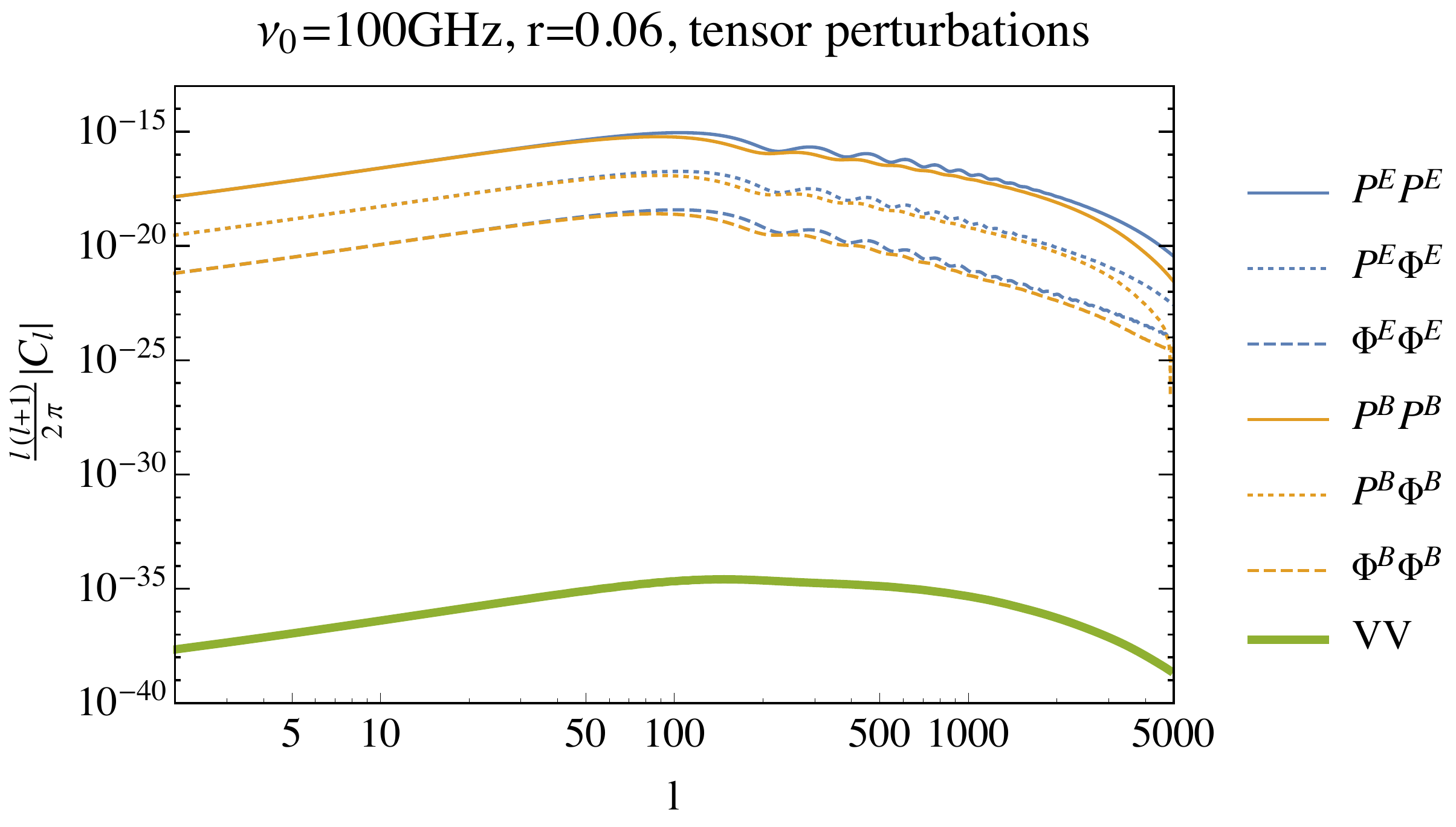}
\caption{The power spectra $C_l$ for tensor perturbations.   The
     blue and green curves are as in
     Fig.~\ref{fig:cl_vv_s}, but we now also have orange curves
     to indicate the analogous quantities for B modes.
     }
\label{fig:cl_vv_t}
\end{figure}

Finally, we provide results of numerical calculations for a
scale-invariant spectrum of primordial gravitational waves,
using \texttt{CLASS}~\cite{Blas:2011rf} to obtain the CMB
polarization transfer functions.
Figs.~\ref{fig:cl_vv_s}, and \ref{fig:cl_vv_t} show
$C_l^{PP}$, $C_l^{P \Phi}$, $C_l^{\Phi\Phi}$, and $C_l^{VV}$
with scalar and tensor perturbations.
We take $\nu_0=100$~GHz for both sets of perturbations.
From Eq.~(\ref{eq:vv_vari}) we find the root-variance of $V$ to be
\begin{align}
\sqrt{\vev{V^2}}  \sim \begin{cases}
     3\times 10^{-14}  & (\text{for scalar perturbations}), \\
     7\times 10^{-18} \left(\frac{r}{0.06} \right)  & (\text{for
     tensor perturbations}),
  \end{cases}
  \label{eq:mean_v}
\end{align}
or in temperature units,
\begin{align}
\sqrt{\vev{V^2}}  \sim \begin{cases}
     8\times 10^{-14} \,\rm{K} & (\text{for scalar perturbations}), \\
     2\times 10^{-17} \left(\frac{r}{0.06} \right) \,\rm{K} & (\text{for tensor perturbations}).
  \end{cases} 
  \label{eq:mean_v_k}  
\end{align}
From Eqs.~(\ref{eq:mean_v}) and (\ref{eq:mean_v_k}), we can see that the circular polarization induced by tensor perturbations through the photon-photon scattering is much larger than that induced through the photon-graviton scattering~\cite{Bartolo:2018igk}.

%%%%%%%%%%%%%%%%%%%%%%%%%%%%%%
\section{Conclusions}
\label{sec:conclusions}
%%%%%%%%%%%%%%%%%%%%%%%%%%%%%%

We have used the TAM formalism to discuss the CMB circular
polarization induced at second order in the
primordial-perturbation amplitude, by general primordial perturbations,
including vector and tensor perturbations in addition to scalar
perturbations.  To make the discussion concrete, we have assumed
the standard cosmology and considered only the dominant
contribution---from photon-photon scattering---to Faraday
conversion.  We performed numerical calculations 
of the power spectra for circular polarization and find
root-variances of $\sqrt{\vev{V^2}}\sim 3\times 10^{-14}$ for scalar
perturbations, and $\sqrt{\vev{V^2}}\sim 7\times
10^{-18}\,(r/0.06)$ for tensor perturbations.  The derived formulas can be
applied to the other source terms discussed in
Ref.~\cite{Montero-Camacho:2018vgs} such as spin polarization
of neutral hydrogen atoms and the non-linear interactions
induced by bounded or free electrons, although these are
expected to be subdominant to the photon-photon process
considered here.

Before closing, we note that it follows from Eq.~(\ref{eqn:vlm})
that the monopole $V_{l=0,m=0}=0$ if there are no primordial vector
or tensor modes (and thus no B modes).  In other words, there
will be circular-polarization fluctuations, but the mean value
of the circular polarization, averaged over the entire sky, will
be zero.  We also note that here we have assumed that parity is
conserved and thus that there are no correlations between the TE
and TB TAM modes.  An accompanying \cite{Inomata:2018rin} paper shows
that the TE/TB cross-correlations that arise if parity is broken
may allow for a parity-breaking uniform (averaged over all
directions) circular polarization $V_{l=0,m=0}$.  The paper also
shows that a uniform circular polarization may arise even in the
absence of parity-breaking physics through a realization of a
gravitational-wave field that spontaneously breaks parity.

\acknowledgments 

We thank Paulo Montero-Camacho for useful comments.
K.I. is supported by World Premier International Research Center
Initiative (WPI Initiative), MEXT, Japan,
Advanced Leading Graduate Course for Photon Science, 
and JSPS Research Fellowship for Young Scientists, 
and thanks Johns Hopkins University for hospitality.  This work
was supported at Johns Hopkins in part by NASA Grant
No. NNX17AK38G, NSF Grant No.\ 1818899, and the Simons
Foundation.

\appendix
%\def\thesection{Appendix \Alph{section}}
%%%%%%%%%%%%%%%%%%%%%%%%%%%%%%
\section{Wigner rotation matrices}
\label{app:useful_f}
%%%%%%%%%%%%%%%%%%%%%%%%%%%%%%

Here we review some useful properties of the Wigner rotation
matrices, using the notation~\cite{wigner2012group},
\begin{align}
\label{eq:wigner_d_def}
D^l_{m',m}(\alpha, \beta, \gamma) =& \sum_{s=\text{max}[0,m-m']}^{\text{min}[l+m,l-m']} (-1)^s \frac{\sqrt{(l+m)!(l-m)!(l+m')!(l-m')!}}{(l-m'-s)!(l+m-s)! s! (s+m'-m)!}  \nonumber \\
& \times \ee^{i m' \alpha} \left( \cos \frac{\beta}{2} \right)^{2l+m-m'-2s} \left(\sin \frac{\beta}{2} \right)^{2s+m'-m} \ee^{i m \gamma}.
\end{align}
The relations between the spherical harmonics in the
line-of-sight frame $(\theta', \phi')$ and the wave-vector frame
$(\theta, \phi)$ is given by~\cite{wigner2012group}
\begin{align}
Y^*_{lm}(\theta', \phi') = \sum_k D^{l}_{mk}(\pi - \phi_{\bm k}, \theta_{\bm k}, 0) Y^*_{lk}(\theta,\phi).
\end{align}
Since the coefficients are given by $a^A_{lm} = \int \dd
\hat{\bm n}\, A(\hat{\bm n})  Y^*_{lm}(\hat{\bm n})$, the
relation between the coefficients of spherical harmonics in the
two frames is
\begin{align}
a^E_{lm}(\bm k, \eta) = \sum_k D^{l}_{mk}(\pi - \phi_{\bm k}, \theta_{\bm k}, 0) \bar{a}^E_{lk}(\bm k, \eta).
\end{align}

%%%%%%%%%%%%%%%%%%%%%%%%%%%%%%
\section{Photon-photon scattering}
\label{app:p_p_scat}
%%%%%%%%%%%%%%%%%%%%%%%%%%%%%%

Here we derive Eqs.~(\ref{eq:nq_def}) and (\ref{eq:nu_def}).
For photons with energies far smaller than the electron
rest-mass energy, the effective Lagrangian for the
electromagnetic field can be approximated as the
Euler-Heisenberg
Lagrangian~\cite{Heisenberg:1935qt}:\footnote{In this paper, we
assume the Standard Model. If we assume new particles, such as
axion-like particles, the coefficients in the Lagrangian could
be changed~\cite{Zavattini:2012zs}.}
\begin{align}
\mathcal L \simeq \frac{1}{2\mu_0} \left( \frac{  \bm{E} \cdot \bm{E}  }{c^2} -  \bm{B} \cdot \bm{B}  \right) + \frac{A_e}{\mu_0} \left[ \left( \frac{ \bm{E} \cdot \bm{E} }{c^2} -  \bm{B} \cdot \bm{B}  \right)^2 + 7 \left( \frac{ \bm{E} \cdot \bm{B} }{c} \right)^2 \right].
\label{eq:e_h_lag}
\end{align}
By using the constitutive relations $\bm{D} = \partial \mathcal L / \partial \bm E$ and $\bm{H} = -\partial \mathcal L / \partial \bm B$, we obtain
\begin{align}
\label{eq:d_express}
\bm D &= \epsilon_0 \bm E + \epsilon_0 A_e \left[ 4 \left( \frac{  \bm{E} \cdot \bm{E}  }{c^2} -  \bm{B} \cdot \bm{B}  \right) \bm E + 14(\bm E \cdot \bm B) \bm B \right], \\
\bm H &= \frac{\bm B}{\mu_0} + \frac{A_e}{\mu_0} \left[ 4 \left( \frac{  \bm{E} \cdot \bm{E}  }{c^2} -  \bm{B} \cdot \bm{B}  \right) \bm B - 14 \frac{(\bm E \cdot \bm B)}{c^2} \bm E \right],
\end{align}
where $\bm D$ is the electric-displacement vector and $\bm H$ is
the magnetic-intensity vector.  To consider the interaction
between the propagating photon and background radiation, we
write the electric and magnetic fields as
\begin{align}
\bm E = \bm E_A \ee^{i(\bm k \cdot \bm x - \omega t)} +  \bm E_A^* \ee^{-i(\bm k \cdot \bm x - \omega t)} + \bm E_B(\bm x,t), \qquad
\bm B = \bm B_A \ee^{i(\bm k \cdot \bm x - \omega t)} +  \bm B_A^* \ee^{-i(\bm k \cdot \bm x - \omega t)} + \bm B_B(\bm x,t),
\label{eq:b_separate}
\end{align}
where $\bm E_A$ and $\bm B_A$ are the electromagnetic fields
associated with the propagating photon and $\bm E_B$ and $\bm
B_B$ are those associated with the background radiation.
We assume $\bm B^A_i = \epsilon_{ijk} \hat k^k \bm E_A^j$, where
$\epsilon_{ijk}$ is the Levi-Civita symbol.  From
Eqs.~(\ref{eq:d_express})--(\ref{eq:b_separate}), we find that
\begin{align}
D_i &\simeq  \ee^{i(\bm k \cdot \bm x - \omega t)} \epsilon_0 \left( \delta_{ij} + A_e \left[ 4 \left( \frac{  \vev{\bm{E_B} \cdot \bm{E_B}}  }{c^2} -  \vev{\bm{B_B} \cdot \bm{B_B}}  \right) \delta_{ij} +  8 \frac{\vev{E^B_i E^B_j}}{c^2} +  14 \vev{B^B_i B^B_j}  \right] \right) E_A^j + \cdots , \\
H_i &\simeq  \ee^{i(\bm k \cdot \bm x - \omega t)}
\frac{1}{\mu_0} \left( \delta_{ij} - (-1) A_e \left[ 4 \left(
\frac{  \vev{\bm{E_B} \cdot \bm{E_B}}  }{c^2} -  \vev{\bm{B_B}
\cdot \bm{B_B}}  \right) \delta_{ij} -  8 \vev{B^B_i B^B_j} -
14 \frac{\vev{E^B_i E^B_j}}{c^2}  \right] \right) \epsilon^{jkl}
\hat k_k E^A_l + \cdots ,
\end{align}
where we explicitly write only the terms proportional to
$\ee^{i(\bm k \cdot \bm x - \omega t)}$, and $\vev{\cdots}$
means the expectation value of the stochastic background
radiation. Then, we derive 
\begin{align}
\label{eq:chi_e}
\chi_{e,ij} &\simeq   A_e \left[ 4 \left( \frac{  \vev{\bm{E_B} \cdot \bm{E_B}}  }{c^2} -  \vev{\bm{B_B} \cdot \bm{B_B}}  \right) \delta_{ij} 
+  8 \frac{\vev{E^B_i E^B_j}}{c^2} 
+  14 \vev{B^B_i B^B_j} \right] , \\
\label{eq:chi_m}
\chi_{m,ij} &\simeq  - A_e \left[ 4 \left( \frac{  \vev{\bm{E_B} \cdot \bm{E_B}}  }{c^2} -  \vev{\bm{B_B} \cdot \bm{B_B}}  \right) \delta^{kl} -  8 \vev{B_B^k B_B^l} -  14 \frac{\vev{E_B^k E_B^l}}{c^2} \right]
 \epsilon_{k m i} \hat k^m \epsilon_{l n j} \hat k^n.
\end{align}
From Eqs.~(\ref{eq:chi_e}) and (\ref{eq:chi_m}), we obtain
\begin{align}
\label{eq:n_xx_n_yy}
n_{xx} (\bm x,t) -n_{yy} (\bm x,t) &= \frac{1}{2} \left( \chi_{e,xx}
+ \chi_{m,xx} - \chi_{e,yy} - \chi_{m,yy} \right) = 3A_e \left(
\vev{B^B_x B^B_x} - \vev{B^B_y B^B_y} - \frac{1}{c^2} \left(
\vev{E^B_x E^B_x} - \vev{E^B_y E^B_y} \right) \right) \\
\label{eq:n_xy}
n_{xy} (\bm x,t) &= \frac{1}{2} \left( \chi_{e,xy} + \chi_{m,xy}
\right) = 3A_e \left( \vev{B^B_x B^B_y} - \frac{1}{c^2}
\vev{E^B_x E^B_y}  \right).
\end{align}
Here, we expand the background electric and magnetic field with
creation and annihilation operators as
\begin{align}
\label{eq:e_back_ex}
E^B_i(\bm x,t) &= i \int \frac{\dd^3 p}{(2\pi)^3} \sum_{\alpha = x, y}  \sqrt{\frac{U_p}{2 \epsilon_0}} \hat{\bm \ee}^\alpha_i 
\left( a_\alpha (\bm p) \ee^{i(\bm p \cdot \bm x - \omega t)} - a_\alpha^\dagger (\bm p) \ee^{-i(\bm p \cdot \bm x - \omega t)} \right),  \\
\label{eq:b_back_ex}
B^B_i(\bm x,t) &= i \int \frac{\dd^3 p}{(2\pi)^3} \sum_{\alpha = x, y}  \sqrt{\frac{U_p}{2 \epsilon_0 c^2}} (\hat{\bm p} \times \hat{\bm \ee}^\alpha_i )
\left( a_\alpha (\bm p) \ee^{i(\bm p \cdot \bm x - \omega t)} - a_\alpha^\dagger (\bm p) \ee^{-i(\bm p \cdot \bm x - \omega t)} \right),
\end{align}
where we define the basis vectors as $\hat{\bm \ee}^x \equiv
\hat{\bm \theta}$ and $\hat{\bm \ee}^y \equiv \hat{\bm \phi}$.  
The expectation value of photon number density is described with a phase-space density matrix as
\begin{align}
\vev{a_\alpha^\dagger (\bm p)\, a_\beta (\bm p')}_{\bm x,t} = (2\pi)^3 \delta( \bm p - \bm p') f_{\alpha \beta} (\bm p, \bm x, t),
\end{align}
where the subscript $\bm x$ and $t$ indicate the space-time point in which we consider the expectation value, and from Ref.~\cite{Kosowsky:1994cy},
\begin{align}
f_{\alpha \beta} (\bm p, \bm x, t) &= \left(
    \begin{array}{cc}
     f_I(\bm p, \bm x, t) + f_Q(\bm p, \bm x, t) & f_U(\bm p, \bm x, t) - i f_V(\bm p, \bm x, t)  \\
     f_U(\bm p, \bm x, t) + i f_V(\bm p, \bm x, t) & f_I(\bm p, \bm x, t) - f_Q(\bm p, \bm x, t) \\
    \end{array}
  \right).
\end{align}
Substituting Eqs.~(\ref{eq:e_back_ex}) and (\ref{eq:b_back_ex})
into Eqs.~(\ref{eq:n_xx_n_yy}) and (\ref{eq:n_xy}), we obtain
\begin{align}
\label{eq:n_xx_n_yy_2}
n_Q(\bm x, t) &= \frac{1}{2} \left( n_{xx} (\bm x, t) -n_{yy} (\bm x, t) \right) \nonumber\\
&=  - \frac{3A_e}{2 \epsilon_0 c^2} \sqrt{\frac{\pi}{5}} \int \frac{U_p}{2} p^2 \dd p \int \dd^2 \hat{\bm p} \left[ 4 f_Q(\bm p, \bm x, t) ( 1+\cos^2\theta_{\bm p} ) \cos 2\phi 
- 8 f_U(\bm p, \bm x, t)  \cos\theta_{\bm p} \sin2\phi_{\bm p} \right] \nonumber \\
&=  - \frac{24A_e}{\epsilon_0 c^2} \sqrt{\frac{\pi}{5}} \int \frac{U_p}{2} p^2 \dd p \int \dd^2 \hat{\bm p} \left[ f_Q(\bm p, \bm x, t)\, \text{Re}\{ _2Y_{22}(\hat {\bm p}) + _2Y_{2,-2}(\hat {\bm p}) \} 
+ f_U(\bm p, \bm x, t)  \, \text{Im} \{ _2Y_{22}(\hat {\bm p}) + _2Y_{2,-2}(\hat {\bm p}) \}  \right] \nonumber \\
&= 48 \sqrt{\frac{\pi}{5}} A_e \mu_0 a_\text{rad}
T_{\text{CMB}}^4 \text{Re}\, a_{2,-2}^E(\bm x, t),
\end{align}
\begin{align}
\label{eq:n_xy_2}
n_U(\bm x, t) &= \frac{1}{2} (n_{xy} (\bm x, t) + n_{yx} (\bm x, t)) \nonumber \\
&= - \frac{3A_e}{2 \epsilon_0 c^2} \sqrt{\frac{\pi}{5}} \int \frac{U_p}{2} p^2 \dd p \int \dd^2 \hat{\bm p} \left[ 4 f_Q(\bm p, \bm x, t) ( 1+\cos^2\theta_{\bm p} ) \sin 2\phi_{\bm p} + 8 f_U(\bm p, \bm x, t)  \cos\theta_{\bm p} \cos 2\phi_{\bm p} \right] \nonumber \\
&=  - \frac{24A_e}{\epsilon_0 c^2} \sqrt{\frac{\pi}{5}} \int
\frac{U_p}{2} p^2 \dd p \int \dd^2 \hat{\bm p} \left[ f_Q(\bm p,
\bm x, t)\, \text{Im}\{ _2Y_{22}(\hat {\bm p}) - _2Y_{2,-2}(\hat
{\bm p}) \} - f_U(\bm p, \bm x, t)  \, \text{Re} \{ _2Y_{22}(\hat
{\bm p}) - _2Y_{2,-2}(\hat {\bm p}) \}  \right] \nonumber \\
&= 48 \sqrt{\frac{\pi}{5}} A_e \mu_0 a_\text{rad} T_{\text{CMB}}^4 \text{Im}\, a_{2,-2}^E(\bm x, t),
\end{align}
where $\theta_{\bm p}$ and $\phi_{\bm p}$ are the polar and
azimuthal angles of $\hat {\bm p}$ in the line-of-sight frame, and
we have used \cite{Lin:2004xy},
\begin{align}
f_Q(\bm p, \bm x, t) &= Q(\hat{\bm p}, \bm x, t) (-p \,\partial
f^{(0)}/ \partial p),  \qquad f_U(\bm p, \bm x, t) = U(\hat{\bm p}, \bm x, t) (-p \,\partial f^{(0)}/ \partial p), \\
Q(\hat{\bm p}, \bm x, t) &= \frac{1}{2}  \sum_{l,m} \left(
a_{2,lm}(\bm x, t) \,_2Y_{lm}(\hat{\bm p}) + a_{-2,lm}(\bm x, t)
\,_{-2} Y_{lm}(\hat{\bm p})  \right),  \\
U(\hat{\bm p}, \bm x, t) &= \frac{1}{2i}  \sum_{l,m} \left( a_{2,lm}(\bm x, t) \,_2Y_{lm}(\hat{\bm p}) - a_{-2,lm}(\bm x, t)  \,_{-2} Y_{lm}(\hat{\bm p})  \right), \\
_sY_{lm}(\hat {\bm p}) &= _{-s}Y_{l,-m}^*(\hat {\bm p}),
\qquad a^E_{lm}(\bm x, t) = - \frac{1}{2} (a_{2,lm}(\bm x, t) + a_{-2,lm}(\bm x, t)).
\end{align}
If we take the conformal space-time, Eqs.~(\ref{eq:n_xx_n_yy_2}) and (\ref{eq:n_xy_2}) correspond to Eqs.~(\ref{eq:nq_def}) and (\ref{eq:nu_def}).

%%%%%%%%%%%%%%%%%%%%%%%%%%%%%%%%%
%%%%%%%%%%% References %%%%%%%%%%%
%%%%%%%%%%%%%%%%%%%%%%%%%%%%%%%%%

\end{document}